\documentclass[11pt]{article}
\usepackage{epsf,psfrag}
\usepackage{graphicx}
\usepackage{amsmath}
\usepackage{amsthm}
\usepackage{amsfonts}
\usepackage{amssymb}
\catcode`\@=11
\def \be  {\begin{equation}}
\def \ee  {\end{equation}}
\def \ba  {\begin{eqnarray}}
\def \ea  {\end{eqnarray}}
\def \baa {\begin{eqnarray*}}
\def \eaa {\end{eqnarray*}}
\def \bb  {\begin {thebibliography} }
\def \eb  {\end{thebibliography}}
\def \lab #1 {\label{#1}}

\def \matrix #1 {\left(\begin{array}{cc} #1 \end{array}\right)}

\def \tr {\mathop{\rm tr}\nolimits}

\newcommand \widebar [1] {\overline{#1}}

\newcommand{\as}{\ifmmode\alpha_{\rm s}\else{$\alpha_{\rm s}$}\fi}
\newcommand{\asbar}{\ifmmode\bar{\alpha}_{\rm s}\else{$\bar{\alpha}_{\rm s}$}\fi}

\newcommand{\CR}{{\mathcal R}}
\newcommand{\CD}{{\mathcal D}}
\newcommand{\VV}{{\mathbb V}}
\newcommand{\CV}{{\mathcal V}}

\newcommand{\WW}{{\mathbb W}}
\newcommand{\CW}{{\mathcal W}}

\newcommand{\TB}{{\mathbb T}}
\newcommand{\CT}{{\mathcal T}}

\newcommand{\TT}{{\mathsf T}}
\newcommand{\CL}{{\mathcal L}}

\newcommand{\bs}{\boldsymbol{\sigma}}
\newcommand{\br}{\boldsymbol{\rho}}

\font\cmss=cmss12 
\def\inbar{\,\vrule height1.5ex width.4pt depth0pt}
\def\IC{\relax\hbox{$\inbar\kern-.3em{\rm C}$}}
\def\IZ{\relax{\hbox{\cmss Z\kern-.4em Z}}}
\def\IR{{\hbox{{\rm I}\kern-.2em\hbox{\rm R}}}}

\def\IP{{\hbox{{\rm I}\kern-.2em\hbox{\rm P}}}}
\def\II{\hbox{{1}\kern-.25em\hbox{l}}}

\def\numberbysection{\@addtoreset{equation}{section}
                     \def\theequation{\thesection\arabic{equation}}}
\numberbysection

\newbox\lett\newdimen\lheight\newdimen\lwidth
\def\ontop#1#2{\setbox\lett=\hbox{#2}\lheight\ht\lett
\multiply\lheight by 12 \divide\lheight by 10\relax%
\lwidth\wd\lett \multiply\lwidth by 8 \divide\lwidth by 10\relax #2\kern-\lwidth%
\raise\lheight\hbox{{$\scriptstyle #1$}}\kern.1ex}

\numberwithin{equation}{section}

\begin{document}

\begin{titlepage}

\vskip3cm
\begin{center}
  { \bf \Large Baxter operators
for the quantum $\boldsymbol{sl(3)}$ invariant spin chain }

\def\thefootnote{\fnsymbol{footnote}}%
\vspace{1cm}
{\sc S. \'{E}. Derkachov}${}^1$
~and~ {\sc A. N.~Manashov}${}^2$\footnote{ Permanent
address:\ Department of Theoretical Physics,  Sankt-Petersburg State University,
St.-Petersburg, Russia}
\\[0.5cm]

\vspace*{0.1cm} ${}^1$ {\it
St.Petersburg Department of Steklov
Mathematical Institute of Russian Academy of Sciences,
Fontanka 27, 191023 St.Petersburg, Russia.
                       } \\[0.2cm]
\vspace*{0.1cm} ${}^2$
 {\it Institute for Theoretical Physics, University of  Regensburg,\\
D-93040 Regensburg, Germany}

\vskip2cm

{\bf Abstract:\\[10pt]} \parbox[t]{15cm}{
The noncompact homogeneous $sl(3)$ invariant spin chains are considered.
We show that the transfer matrix with generic auxiliary space
is factorized into the product of three $sl(3)$ invariant commuting
operators.
These operators satisfy the finite difference equations in the spectral
parameters which follow from the structure of the reducible $sl(3)$ modules.
}
\vskip1cm

\end{center}

\end{titlepage}


\section{Introduction}\label{sect:intro}
In this paper we address the problem of constructing  the Baxter $Q-$operators for
the integrable $sl(3)$ invariant noncompact spin magnet. As it is well known the spin magnets
can be solved with the help of the Algebraic Bethe
Ansatz (ABA)~\cite{FST,LF}, or its extension to the  symmetry groups of higher rank,
the Nested Bethe Ansatz~\cite{Sut,KR82,KR83}. Alternative approaches for solving spin chain
models are  the method of Baxter $Q-$operators~\cite{Baxter} and
Separation of Variables~\cite{Sklyanin}. The latter, however,
are not  used so widely as the ABA
method. It is in a great extent related to the fact that
there  no regular method exists for  constructing  the Baxter operators or the
representation of Separated variables for the spin chains with symmetry group of the rank
greater than two. On the other hand these methods can be used to study  models
which do not belong  to the range of  applicability  of the ABA,
such as the Toda chain, the $sl(2,C)$ and $sl(2,R)$ noncompact spin magnets,
the modular $XXZ$ magnet.
The Baxter operators for these models and some others were constructed in
Refs.~\cite{PG,V95,BLZ,SD,Pronko,KSS,DKM01,RW,open,KM,Korff1,YNZ,YZ,Bytsko}. For example,
Baxter operators for a different kind of the $sl(2)$ magnets can be
obtained with the help of the Pasquier-Gaudin trick~\cite{PG}. However, the generalization
of  this method to the  higher rank groups  ($sl(N)$) seems to be quite problematic.

In the present paper we develop a method to construct  Baxter
$Q-$operators for the $sl(3)$ spin magnet. This model is well studied -- the
Nested Bethe Ansatz had been developed by Kulish and Reshetikhin~\cite{KR82}, the
Separation of Variables was constructed in the works of
Sklyanin~\cite{Sklyanin-cl,Sklyanin-sl3}. The connection of the Nested Bethe Ansatz with Baxter
equation were investigated by Pronko and Stroganov~\cite{PS}.

The approach presented here generalizes the method developed in~\cite{DM05} for
the $sl(2)$ spin magnet (see also Ref.~\cite{AF} where similar arguments were applied to the analysis
of the $q-$deformed spin chain models).
Our analysis is based  on two main ingredients.
 First of them  is the factorization property of
$\CR-$operator acting on  the tensor product of
two generic $sl(3)$ modules~\cite{SD1}.
We shall show that this property results in the factorization of the
transfer matrices into the product of three operators (Baxter operators). The latter depend on
a spectral parameter and commute with each other.
The second ingredient is the analysis of  properties of the transfer matrices with
a reducible auxiliary space.
We shall show that
transfer matrices with a finite dimensional auxiliary space can be represented by
a certain combinations of  generic transfer matrices. Such a representation is
in one to one correspondence with the decomposition of the reducible $sl(3)$ modules onto
irreducible ones. Together with the factorization of  generic transfer
matrices into the product of three operators it gives rise to a
certain type of self-consistency  equations involving the Baxter operators and the auxiliary
transfer matrices. Thus the structure of the Baxter equation reflects
the structure of the  decomposition of the reducible $sl(3)$ modules onto
irreducible ones.
Although we consider the $sl(3)$ invariant spin chain,
it seems that this method can provide one with a regular method to construct Baxter
operators for   quantum $sl(N)$ invariant spin magnets.

The paper is organized as follows. In Section~\ref{sect:1} we introduce the necessary
notations and describe the model. In Section~\ref{sect:fact} we prove the factorization
of a generic transfer matrix into the product of  Baxter  operators.
In Section~\ref{sect:red-mod} we analyze the structure of  reducible $sl(3)$ modules and
obtain the expression for the auxiliary transfer matrices in terms of the generic ones.
In Section~\ref{sect:baxter} the derivation of the Baxter equation is presented and
Section~\ref{sect:summary} contains  concluding remarks. The appendices contain some
technical details.

\section{Preliminaries}\label{sect:1}

The fundamental object in the theory of  lattice integrable models is the $\mathcal
R-$operator. It is a linear operator which
depends on  a spectral parameter $u$ and  acts on  tensor product of
two $s\ell(3)$ modules (representations of the $s\ell(3)$
algebra). The $\mathcal{R}-$operator satisfies the Yang-Baxter relation (YBR)
\begin{equation}    \label{YB}
\CR_{12}(u)\CR_{13}(u+v)\CR_{23}(v)=\CR_{23}(v)\CR_{13}(u+v)\CR_{12}(u)\,.
\end{equation}
The operators in~\eqref{YB} act on the tensor product of the $sl(3)$ modules,
$\VV_1\otimes\VV_2\otimes\VV_3$, and, as usual, indices
$ik$ indicate that the operator $\CR_{ik}$ acts
nontrivially on the tensor product $\VV_i\otimes\VV_k$. We
shall consider the $sl(3)$ invariant solutions of the YBR.

Throughout the paper we shall use the following realization of the $sl(3)$ module $\VV$.
As the vector space $\mathbb{V}$ we take
the space of polynomials of three complex variables, $\mathbb{C}[x,y,z]$.
The generators of the $sl(3)$ algebra take the form of
differential operators
\begin{subequations}\label{generators}
\begin{align}
&L_{21}=-\partial_{x}\,,\qquad
L_{31}=-\partial_{y}\,,\qquad
L_{32}=-\partial_{z}-x\partial_{y}\,,\\[2mm]
&L_{12}=x^2\partial_{x}+x(y\partial_y-z\partial_{z})+y\partial_{z}+m_1 x\,,\\[2mm]
&L_{13}=y(y\partial_y+z\partial_{z}+
x\partial_{x})-xz^2\partial_{z}+(m_1+m_2)y
-m_2 xz\,,\\[2mm]
&L_{23}=z^2\partial_{z}-y\partial_{x}+m_2 z\,,\\[2mm]
 &H_1=2x\partial_{x}+ y\partial_{y}-z\partial_{z}+m_1\,,\qquad
H_2=2z\partial_{z}+y\partial_{y}- x\partial_{x}+m_2\,.
\end{align}
\end{subequations}
They satisfy the standard $sl(3)$ commutation relations
\begin{align}
[L_{ab} L_{cd}]=\delta_{cb}L_{ad}-\delta_{ad}L_{cb}\,,
\end{align}
where
\begin{align}\label{L-diag}
L_{11}=\frac23 H_1+\frac13 H_2\,,&&L_{22}=\frac13 H_2-\frac13 H_1\,,&& L_{33}=-\frac13
H_1-\frac23 H_2\,.
\end{align}
The module $\mathbb{V}$ is completely determined by the
eigenvalues of the Cartan generators  $H_1, H_2$ on the lowest weight vector $\Phi=1$,
 ($H_1\Phi=m_1\Phi$, $H_2\Phi=m_2\Phi$) and will be denoted as
$\mathbb{V}_{\boldsymbol{m}}$, where $\boldsymbol{m} = (m_1
, m_2)$. Unless neither of numbers $2-m_1-m_2$, $1-m_1$ or $1-m_2$ is a positive integer,
the module $V_{\boldsymbol{m}}$ is
irreducible. The reducible modules will play an  important
role in our analysis and will be discussed in
sect.~\ref{sect:red-mod}.

We shall also use  another notation for the $sl(3)$ modules,
$\mathbb{V}_{\boldsymbol{\sigma}}\equiv\mathbb{V}_{\boldsymbol{m}}$.
Instead of the weights
$m_1,m_2$ one can label a $sl(3)$ module $\mathbb{V}$ by the
three vector
$\boldsymbol{\sigma}=(\sigma_1,\sigma_2,\sigma_3)$, where
\begin{align}\label{mk}
m_1=\sigma_{2}-\sigma_{1}+1\,,&&
m_2=\sigma_{3}-\sigma_{2}+1\,,&&
\sigma_1+\sigma_2+\sigma_3=0
\end{align}
or, explicitly
\begin{align}\label{123}
\sigma_1 = 1-\frac{2 m_1}{3}-\frac{m_2}{3}\,, && \sigma_2 =
\frac{m_1}{3}-\frac{m_2}{3}\,, && \sigma_3 =
-1+\frac{m_1}{3}+\frac{2 m_2}{3}\,.
\end{align}
Convenience of such notation will become clear later.

Provided that the solution of the YBR~\eqref{YB} is known, one can construct the family of
commuting operators -- transfer matrices~\cite{FST,LF}:
\begin{align}\label{TT}
\TT_{\boldsymbol{m}}(u)=\tr \CR_{10}(u)\ldots\CR_{N0}(u)\,.
\end{align}
The trace in Eq.~\eqref{TT} is taken over the auxiliary space $\VV_0\equiv\VV_{\boldsymbol{m}}$
and $ \TT_{{\boldsymbol{m}}}(u)$ acts on the tensor product of the $sl(3)$ modules,
\begin{align}
\mathcal{V}=\VV_1\otimes\VV_2\otimes\ldots\otimes\VV_N\,.
\end{align}
We shall consider  the homogeneous spin chains only, i.e. assume that the quantum spaces
 $\VV_k$,  $k=1,\ldots,N$  have the same
``quantum numbers'', $\VV_k=\VV_{\boldsymbol{n}}$, ($\boldsymbol{n}=(n_1,n_2)$).
By virtue of the YBR relation, the transfer matrices $T_{\boldsymbol{m}}(u)$
commute with each other
for  different values of the spectral parameters and spins of auxiliary space~$\boldsymbol{m}$
\begin{align}\label{Tcom}
[\TT_{\boldsymbol{m}}(u),\TT_{\boldsymbol{m}'}(v)]=0\,.
\end{align}
The above equation implies that the transfer matrices share the common set
of the eigenfunctions.

\section{Factorization}\label{sect:fact}
The $\CR-$operator on the tensor product of two generic $sl(3)$ modules,
$
\VV_{\boldsymbol{n}}\otimes \VV_{\boldsymbol{m}}
\equiv
\VV_{\boldsymbol{\rho}}\otimes \VV_{\boldsymbol{\sigma}},
$
can be obtained as the solution of the $RLL$ relation~\cite{KRS}
\begin{align}\label{RLL}
\CR_{12}(u-v)\, L_1(u)\,L_2(v)=L_2(v)\,L_1(u)\,\CR_{12}(u-v)\,.
\end{align}
Here the operator $\CR_{12}\equiv \CR_{\boldsymbol{n}\boldsymbol{m}}
 \equiv \CR_{\boldsymbol{\rho}\boldsymbol{\sigma}}$ acts on the
tensor product $\VV_{\boldsymbol{\rho}}\otimes \VV_{\boldsymbol{\sigma}}$ and Lax
operators $L_1$ and $L_2$ --- on the tensor products
$\mathbb{C}^3\otimes \VV_{\boldsymbol{\rho}}$ and
$\mathbb{C}^3\otimes \VV_{\boldsymbol{\sigma}}$, respectively.
The Lax operator has the following form
\begin{align}
L(u)=u+\left(
\begin{array}{ccc}
L_{11}& L_{21} &L_{31}\\
L_{12}&L_{22}& L_{32}\\
L_{13}&L_{23} &L_{33}
\end{array}
\right)
\end{align}
with the  generators $L_{ik}$
defined in Eqs.~\eqref{generators},~\eqref{L-diag}. It depends on  three
parameters -- the spectral parameter $u$ and two spins $m_1,m_2$. It is convenient to
define the following independent variables
\begin{align}\label{usigma}
u_k=u-1-\sigma_k, \qquad k=1,2,3\,.
\end{align}
The parameters $u_1,u_2,u_3$ define
unambiguously
the parameters $u,m_1,m_2$, and, as a consequence, the Lax operator
$L(u)=L(u_1,u_2,u_3)=L(\boldsymbol{u})$.
In the Ref.~\cite{SD1} it was suggested to look for the solution of Eq.~\eqref{RLL}
in the factorized form
\begin{equation}\label{R123}
\CR_{12}(u)= \mathrm{P}_{12} \CR_{1}\CR_2\CR_3\,,
\end{equation}
where $\mathrm{P}_{12}$ is the permutation operator. The
defining equations for the $\CR_k$ operators are
\begin{align}   \label{123LL} \CR_k\,
L_1(\boldsymbol{u}) \,L_2(\boldsymbol{v})=
L_1(\boldsymbol{u}_{k}) \,L_2(\boldsymbol{v}_{k})\,\CR_k\,.
\end{align}
Here the vectors $\boldsymbol{u}_{k}$,
$\boldsymbol{v}_{k}$ have the interchanged components $u_k$
and $v_k$ in comparison with $\boldsymbol{u}$ and
$\boldsymbol{v}$, for example
$$
\boldsymbol{u}_{1} =(v_1,u_2,u_3)\,,\ \ \ \ \ \boldsymbol{v}_{1}
=(u_1,v_2,v_3)\,,
$$
where $v_k=v-1-\rho_k$.
In other words, the action of the operator   $\CR_k$
results in the interchange of the arguments
$u_k$ and $v_k$ in the Lax operators,
$$
\CR_1 L_1(u_1,u_2,u_3)L_2(v_1,v_2,v_3)=
L_1(v_1,u_2,u_3)L_2(u_1,v_2,v_3)\CR_1\,,
$$
and so on. It is easy to see that the YB relation~\eqref{RLL} holds for the $\CR-$ operator
~\eqref{R123} provided that the operators $\CR_k$ satisfy Eqs.~\eqref{123LL}. Indeed,
 using repeatedly Eq.~\eqref{123LL} for the operators $\CR_3$, $\CR_2$ and $\CR_1$ one derives
$$
\CR_{1}\CR_2\CR_3 L_1(u_1,u_2,u_3)L_2(v_1,v_2,v_3)=
L_1(v_1,v_2,v_3)L_2(u_1,u_2,u_3)\CR_{1}\CR_2\CR_3\,.
$$
Taking into account that $P_{12}
L_1(\boldsymbol{v})L_2(\boldsymbol{u})P_{12}^{-1}=
L_2(\boldsymbol{v})L_1(\boldsymbol{u})$ one gets the
necessary result. Since the vector spaces $\VV_1$ and
$\VV_2$ are isomorphic, the operator $P_{12}$ is well
defined on the tensor product $\VV_1\otimes\VV_2$. The
operators $\CR_k$ which satisfy Eq.~\eqref{123LL}
were constructed in \cite{SD1}.
Each operator $\CR_k$ depends on the subset of the spectral parameters, $\boldsymbol{u}$ and
$\boldsymbol{v}$, only,
\begin{align}\label{CRK}
\CR_1=\CR_1(u_1|v_1,v_2,v_3)\,,&&
\CR_2=\CR_2(u_1,u_2|v_2,v_3)\,,&&
\CR_3=\CR_3(u_1,u_2,u_3|v_3)\,
\end{align}
and is invariant under a simultaneous shift of all spectral parameters
$u_i\to u_i+a$,
$v_i\to v_i+a$.
We give the explicit
expressions for the operators $\CR_k$ in
Appendix~\ref{app:operators}. Here we want to discuss briefly their
properties.  It follows from the relation~\eqref{123LL}  that
operators $\CR_k$ are $sl(3)$ covariant, i.e.
\begin{align}
\CR_k: \VV_{\boldsymbol{n}}\otimes
\VV_{\boldsymbol{m}}\mapsto \VV_{\boldsymbol{n}_{k}}\otimes
\VV_{\boldsymbol{m}_{k}}\,,
\end{align}
where
\begin{subequations}\label{spins}
\begin{align}
\boldsymbol{n}_{1}&=(n_1-\lambda_1,n_2)\,,&
\boldsymbol{m}_{1}&=(m_1+\lambda_1,m_2)\,,&
&\lambda_1=u_1-v_1\,,\\[2mm]
\boldsymbol{n}_{2}&=(n_1+\lambda_2,n_2-\lambda_2)\,,&
\boldsymbol{m}_{2}&=(m_1-\lambda_2,m_2+\lambda_2)\,,&
&\lambda_2=u_2-v_2\,,\\[2mm]
\boldsymbol{n}_{3}&=(n_1,n_2+\lambda_3)\,,&
\boldsymbol{m}_{3}&=(m_1,m_2-\lambda_3)\,,&
&\lambda_3=u_3-v_3\,.
\end{align}
\end{subequations}
The action of the $\CR-$operator~\eqref{R123} on the tensor product
$ \VV_{\boldsymbol{n}}\otimes\VV_{\boldsymbol{m}}$ results in the following chain
of transformations
\begin{align}
\CR_{12}(u):& \VV_{n_1,n_2}\otimes\VV_{m_1,m_2}\xrightarrow{\ \CR_3\  }
\VV_{n_1,n_2+\lambda_3}\otimes\VV_{m_1,m_2-\lambda_3}\xrightarrow{\ \CR_2\  }\nonumber\\[2mm]
&\VV_{n_1+\lambda_2,m_2}\otimes\VV_{m_1-\lambda_2,n_2}\xrightarrow{\ \CR_1\  }
\VV_{m_1,m_2}\otimes\VV_{n_1,n_2}\xrightarrow{\ P_{12}\  }\VV_{n_1,n_2}\otimes\VV_{m_1,m_2}\,,
\end{align}
where we have taken into account the relations~\eqref{usigma} and \eqref{mk}.

The operators $\CR_k$ are completely determined by the
spins of the spaces $\VV_{\boldsymbol{n}}\otimes
\VV_{\boldsymbol{m}}$ that they  act on and the spectral
parameter~$\lambda_k$, namely
\begin{align}\label{Rhalf}
\CR_1=\CR_1(\lambda_1|m_1,m_2)\,,&&\CR_2=\CR_2(\lambda_2|n_1,m_2)
\,,&&\CR_3=\CR_3(\lambda_3|n_1,n_2)\,.
\end{align}
One sees that the operator $\CR_1$ depends on the spins of the second space,
$\VV_{\boldsymbol{m}}$, the operator $\CR_3$ -- on the spins of the first
space, $\VV_{\boldsymbol{n}}$, and $\CR_2$ -- on the spins $n_1$ and $m_2$.

In what follows we shall often display the dependence of
the operators $\CR_k$ on the spectral parameter only,
$\CR_k=\CR_k(\lambda_k)$, always implying that the others
parameters are determined by the tensor properties of the
tensor product $\VV_{\boldsymbol{n}}\otimes
\VV_{\boldsymbol{m}}$.

Note also that for the spectral parameters $u_k=v_k$
($\lambda_k=0$) the operator $\CR_k$ turns into the unit
operator, $\CR_k(0)=\mathbb{I}$. The $\CR-$operator can be
represented as
\begin{align}\label{R6}
\CR_{12}(u-v)=\rho_{k_1k_2k_3}P_{12}\,
\CR_{k_1}(\lambda_{k_1})\,\CR_{k_2}(\lambda_{k_2})
\,\CR_{k_3}(\lambda_{k_3})\,,
\end{align}
where $(k_1,k_2,k_3)$ is the arbitrary permutation of
$(1,2,3)$ and $\lambda_k$ are defined in~\eqref{spins}.
Indeed,  it follows from Eq.~\eqref{123LL} that
$\CR-$matrix~\eqref{R6} satisfies the YB
relation~\eqref{RLL} for any permutation $S$ and, therefore,
can differ  from \eqref{R123} by a normalization coefficient ($\rho_{k_1k_2k_3}$) only.
 To find the latter it is sufficient to compare the
eigenvalues of the $\CR-$ operator  for  the lowest weight
vector, $\Phi=1$.

Let us denote by $\CR_k^{(ij)}$ the operator $\CR_k$
acting on the tensor product $\VV_i\otimes \VV_j$.
One can easily check that for $i\neq k$ the operator
$\CR_{i}^{(12)}(u_i-v_i)\CR_k^{(23)}(v_k-w_k)$ results
in the same permutation of
the arguments
in the product of the Lax operators,
 $L_1(\boldsymbol{u}) L_2(\boldsymbol{v}) L_3(\boldsymbol{w})$, as the
operator \mbox{$\CR_k^{(23)}(v_k-w_k)\CR_{i}^{(12)}(u_i-v_i)$}.
Thus one concludes that $
\CR_{i}^{(12)}(\lambda)\CR_k^{(23)}(\mu)\sim
\CR_k^{(23)}(\mu)\CR_{i}^{(12)}(\lambda)$.
One can verify  using explicit expressions for the
$\CR_k-$operators~(see Appendix~\ref{app:operators})
that for $i>k$ the coefficient of proportionality is  equal to $1$,  i.e.
\begin{align}\label{RRcomm}
\CR_{i}^{(12)}(\lambda)\CR_k^{(23)}(\mu)=
\CR_k^{(23)}(\mu)\CR_{i}^{(12)}(\lambda).
\end{align}


\subsection{Baxter operators}
\begin{figure}[t]
\psfrag{RR}[cr][cc][1.1]{$\left[\CR_k^{i0}(\lambda_k)\right]^{j_i j_0}_{j'_i j'_0}\qquad=$}
\psfrag{0}[cc][cc][1.1]{$j_0$}
\psfrag{0a}[cc][cc][1.1]{$j'_{0}$}
\psfrag{i}[cc][cc][1.1]{$j_i$}
\psfrag{ia}[cc][cc][1.1]{$j'_{i}$}
\psfrag{R}[cc][cc][1.1]{$\CR_k$}

\centerline{\includegraphics[scale=0.6]{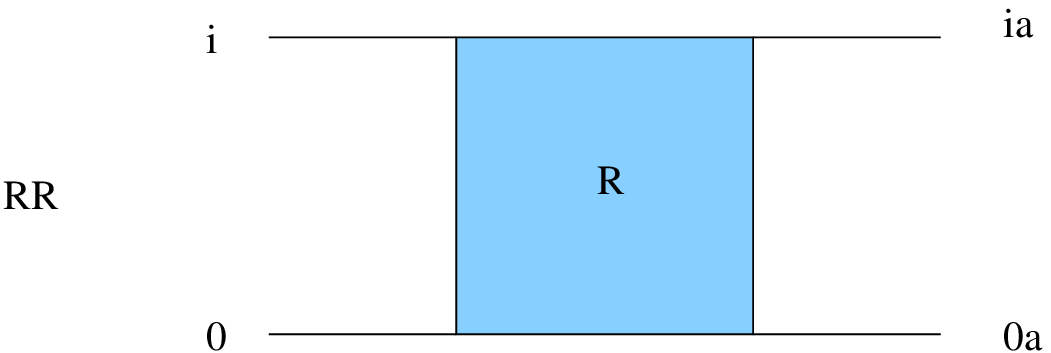}}
\caption[]{The graphical representation of  the matrix of the $\CR_k$ operator.}
\label{fig:Rk}
\end{figure}
Let us remind that, by  construction, the operator $\CR_3(\lambda)$ maps
$\VV_{\boldsymbol{n}}\otimes \VV_{\boldsymbol{m}}\to
\VV_{\boldsymbol{n}'}\otimes \VV_{\boldsymbol{m}'}$. At the same time, this
operator depends on the spins of the first space, $(n_1,n_2)$, only, or, which is the same, it depends
on three
variables $u_1-v_3$, $u_2-v_3$ and $u_3-v_3=\lambda_3$. Therefore, the operator
$\CR_3(u_3-v_3|n_1,n_2)$ satisfies the relation \eqref{123LL} for the arbitrary parameters
$v_1,v_2$, (the arbitrary spins $m_1,m_2$ of the second space).
Having set $v_1=u_1, v_2=u_2$,
($\widetilde{m}_1=n_1, \widetilde{m}_2=n_2+\lambda_3$) one finds out that the operator
$\CR_3(u_3-v_3|n_1,n_2)$ maps $\VV_{\boldsymbol{n}} \otimes\VV_{\widetilde{\boldsymbol{m}}}$ to
$\VV_{\widetilde{\boldsymbol{m}}}\otimes \VV_{\boldsymbol{n}}$.
This implies that the operator $\mathcal{L}^{(3)}(\lambda_3)=P_{12}\CR_3(u_3-v_3|n_1,n_2)$
is the $sl(3)$ invariant operator on the space
$\VV_{\boldsymbol{n}}\otimes \VV_{\widetilde{\boldsymbol{m}}}$ ,
\begin{align}
\mathcal{L}^{(3)}(\lambda_3)(u):& \VV_{n_1,n_2}\otimes\VV_{n_1,n_2+\lambda_3}\xrightarrow{\ \CR_3\  }
\VV_{n_1,n_2+\lambda_3}\otimes\VV_{n_1,n_2}\xrightarrow{\ P_{12}\  }
\VV_{n_1,n_2}\otimes\VV_{n_1,n_2+\lambda_3}\,,
\end{align}
that satisfies the relation
\begin{align}\label{LRL}
\mathcal{L}^{(3)}(\lambda_3) L_1(u_1,u_2,u_3)L_2(u_1,u_2,v_3)=
L_2(u_1,u_2,v_3)L_1(u_1,u_2,u_3)
\mathcal{L}^{(3)}(\lambda_3)\,.
\end{align}
Thus the operator $\mathcal{L}^{(3)}(u)$ has to coincide with the $\CR$ matrix
on the tensor product $\VV_{n_1,n_2}\otimes \VV_{n_1,n_2+u}$. Indeed, noticing
that for $u_1=v_1$ and $u_2=v_2$ the spectral parameters $\lambda_1=\lambda_2=0$,
one gets from Eq.~\eqref{R6}
\begin{align}\label{LR}
\mathcal{L}^{(3)}(u)=\CR_{\boldsymbol{n}
\widetilde{\boldsymbol{m}}}\left(\frac{u}{3}\right)=
\CR_{(n_1,n_2),(n_1,n_2+u)}\left(\frac{u}{3}\right)\,.
\end{align}
(We switched to the standard notation for the spectral parameter, $\lambda\to u$.)
The similar considerations hold  for the operators $\CR_1$, $\CR_2$ as well,
resulting
in the following identification
\begin{align}\label{RL12}
\mathcal{L}^{(1)}(u)&=
\CR_{(n_1,n_2),(n_1-u,n_2)}\left(\frac{u}{3}\right)\,,\\[2mm]
\mathcal{L}^{(2)}(u)&=
\CR_{(n_1,n_2),(n_1+u,n_2-u)}\left(\frac{u}{3}\right)\,,
\end{align}
where $\mathcal{L}^{(k)}(u)=P_{12}\CR_k(u)$.
Since all spaces $\VV_{\boldsymbol{m}}$ and $\VV_{\boldsymbol{m}'}$
are isomorphic to each other (as vector spaces) one concludes that the trace of the product
of the   operators $\mathcal{L}_k$
 is the $sl(3)$ invariant operator on $\VV_1\otimes\ldots\otimes \VV_N$.
For example, it follows from
Eq.~\eqref{LR} that
\begin{align}\label{Tinv}
\tr_{\VV_{\boldsymbol{m}}}\mathcal{L}^{(1)}_{10}(u)\ldots\mathcal{L}^{(1)}_{N0}(u)~=~
\TT
_{(n_1-u,n_2)}\left(\frac{u}{3}\right)\,,
\end{align}
where $\VV_{\boldsymbol{m}}$ is arbitrary and $\VV_i=\VV_{\boldsymbol{n}}$, $i=1,\ldots,N$.

Next we  define three $sl(3)$ invariant operators, $Q_k(u)$, acting
on $\VV_1\otimes\VV_2\otimes\ldots\VV_N$
by
\begin{subequations}\label{Q123}
\begin{align}
Q_1(u+\rho_1)&=
\mathcal{P}^{-1}\tr_{\VV_{\boldsymbol{m}}}\CL_{10}^{(1)}(u)\ldots \CL^{(1)}_{N0}(u)=
\mathcal{P}^{-1}\TT_{(n_1-u,n_2)}\left(\frac{u}{3}\right)
\,,\\[4mm]
Q_2(u+\rho_2)&=
\mathcal{P}^{-1}\tr_{\VV_{\boldsymbol{m}}}\CL_{10}^{(2)}(u)\ldots \CL^{(2)}_{N0}(u)=
\mathcal{P}^{-1}\TT_{(n_1+u,n_2-u)}\left(\frac{u}{3}\right)\,,\\[4mm]
Q_3(u+\rho_3)&=\tr_{\VV_{\boldsymbol{m}}}\CL_{10}^{(3)}(u)\ldots \CL^{(3)}_{N0}(u)=
\TT_{(n_1,n_2+u)}\left(\frac{u}{3}\right)\,.
\end{align}
\end{subequations}
Here the parameters $\rho_k$ specify the quantum space,
\begin{align}
n_1=\rho_{2}-\rho_1+1,& &
n_2=\rho_{3}-\rho_2+1,& &\rho_1+\rho_2+\rho_3=0\,.
\end{align}
The  operator $\mathcal{P}$ is the operator of cyclic permutations
\begin{align}\label{perm}
\mathcal{P}\,\Phi(\boldsymbol{x}_1,\ldots,\boldsymbol{x}_N)=
\Phi(\boldsymbol{x}_N,\boldsymbol{x}_1,\ldots,\boldsymbol{x}_{N-1})\,.
\end{align}
The specific form of the arguments of the Baxter operators $Q_k$ in l.h.s. of Eqs.~\eqref{Q123}
and  presence of the operator $\mathcal{P}$
in the definitions of the operators $Q_{1}$ and $Q_2$ are  matter of convenience.

\begin{figure}[t]
\psfrag{i}[cc][cc][1.1]{$\CR_i$}
\psfrag{k}[cc][cc][1.1]{$\CR_k$}
\psfrag{1}[cc][cc][1.0]{$i_1$}
\psfrag{2}[cc][cc][1.0]{$i_2$}
\psfrag{3}[cc][cc][1.0]{$i_3$}
\psfrag{a}[cc][cc][1.0]{$i'_1$}
\psfrag{b}[cc][cc][1.0]{$i'_2$}
\psfrag{c}[cc][cc][1.0]{$i'_3$}

\psfrag{=}[cc][cc][1.4]{$=$}
\centerline{\includegraphics[scale=0.6]{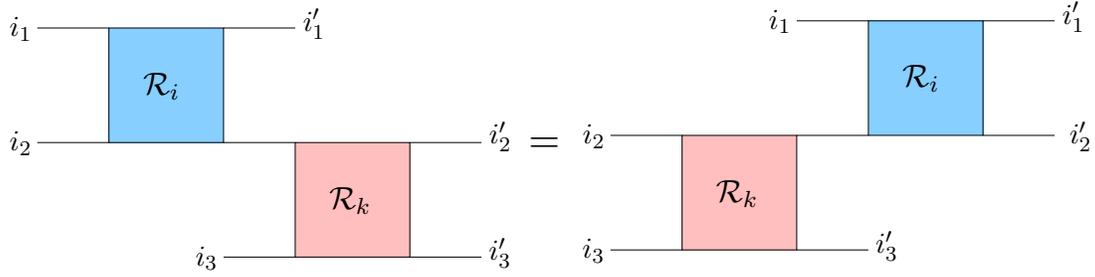}}
\caption[]{The graphical representation of the permutation identity~\eqref{RRcomm}.} \label{fig:RR}
\end{figure}

Having put $u=0$ in Eqs.~\eqref{Q123} one finds
$$
Q_1(\rho_1)=Q_2(\rho_2)=\mathbb{I}\,,\qquad Q_3(\rho_3)=\mathcal{P}\,.
$$
In the  $\boldsymbol{\rho}$ notations
the expressions for the $Q$ operators take a more symmetric form.
\begin{subequations}\label{Qs}
\begin{align}
Q_1(u+\rho_1)&=\mathcal{P}^{-1}
\TT_{\rho_1+2\alpha,\rho_2-\alpha,\rho_3-\alpha}
(\alpha)\Bigl|_{\alpha=u/3}\,,\\[2mm]
Q_2(u+\rho_2)&=\mathcal{P}^{-1}
\TT_{\rho_1-\alpha,\rho_2+2\alpha,\rho_3-\alpha}
(\alpha)\Bigl|_{\alpha=u/3}\,,\\[2mm]
Q_3(u+\rho_3)&=\TT_{\rho_1-\alpha,\rho_2-\alpha,\rho_3+2\alpha}
(\alpha)\Bigl|_{\alpha=u/3}\,.
\end{align}
\end{subequations}
It follows from the commutativity of the transfer matrices~\eqref{Tcom}
that the operators $Q_k(u)$ commute with each
other
\begin{align}
[Q_k(u),Q_j(v)]=0.
\end{align}
Moreover, we shall prove
that the transfer matrix is factorized into the product of these operators
\begin{align}\label{T3Q}
\TT_{(m_1,m_2)}(u)\equiv
\TT_{(\sigma_1,\sigma_2,\sigma_3)}(u)
=Q_1(u+\sigma_1)\,Q_2(u+\sigma_2)\,Q_3(u+\sigma_3)\,.
\end{align}
\begin{figure}[t]
\psfrag{1}[cc][cc][1.1]{$i_1$}
\psfrag{2}[cc][cc][1.1]{$i_2$}
\psfrag{3}[cc][cc][1.1]{$i_N$}
\psfrag{1a}[cc][cc][1.1]{$i'_1$}
\psfrag{2a}[cc][cc][1.1]{$i'_2$}
\psfrag{3a}[cc][cc][1.1]{$i'_N$}

\psfrag{a}[cc][cc][1.0]{$\CR_{12}$}
\psfrag{b}[cc][cc][1.0]{$\CR_3$}

\psfrag{=}[cc][cc][1.4]{$=$}
\centerline{\includegraphics[scale=0.9]{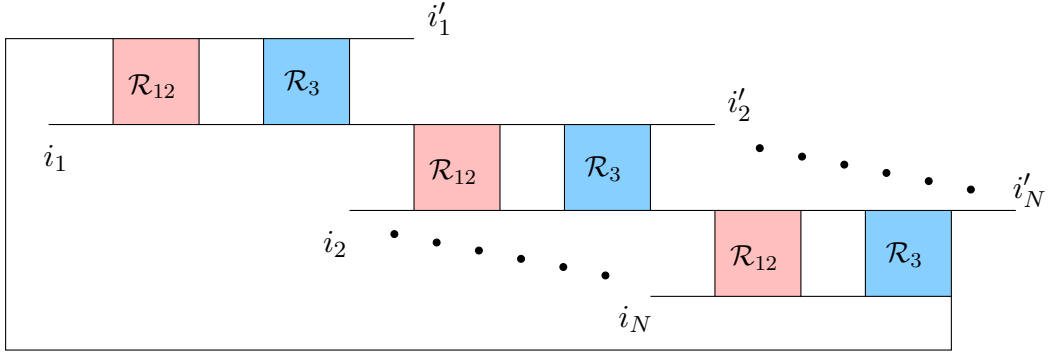}}
\caption[]{The graphical representation for the transfer matrix~\eqref{TT}.
The boxes with the label $\CR_{12}$  denote the matrix the operator $\CR_{12}\equiv\CR_1\CR_2$.}
\label{fig:ladder}
\end{figure}

The proof is based on the disentangling of the trace~\eqref{TT} with the help of the
commutation relations~\eqref{RRcomm}. Let us choose  some basis in the space
$\VV=\mathbb{C}[x,y,z]$. Then  an arbitrary operator $A:\VV\otimes\VV\mapsto\VV\otimes\VV$
is represented by the matrix $A^{i\:\:j}_{i'j'}$ in the corresponding basis,
$\psi=\sum_{ij}\psi^{ij}(e_i\otimes e_j)$, $(A\psi)^{ij}=\sum_{i'j'}A^{ij}_{i'j'}\psi^{i'j'}$.
To simplify the combinatoric it is convenient to represent the matrix $A^{i\:\:j}_{i'j'}$
by the box with four attached lines corresponding to the indices $(i,j),(i',j')$,
see Fig.~\ref{fig:Rk}.
The line connecting two boxes will imply the summation over the corresponding index.
For example, the product
$(AB)^{i\:\:j}_{i''j''}=\sum_{i'j'}A^{i\:\:j}_{i'j'}B^{i'\:\:j'}_{i''j''} $ is
represented by two boxes which are connected by two lines.
The  commutation relation~\eqref{RRcomm} is represented by the diagram
shown in Fig.~\ref{fig:RR}\,.

\begin{figure}[t]
\psfrag{1}[cc][cc][1.0]{$i_1$}
\psfrag{2}[cc][cc][1.0]{$i_2$}
\psfrag{N}[cc][cc][1.0]{$i_N$}
\psfrag{1a}[cc][cc][1.0]{$i'_1$}
\psfrag{N1a}[cc][lc][1.0]{$i'_{N-1}$}
\psfrag{Na}[cc][cc][1.0]{$i'_N$}
\psfrag{a}[cc][cc][1.1]{$\CR_{12}$}
\psfrag{b}[cc][cc][1.1]{$\CR_3$}

\centerline{\includegraphics[scale=0.8]{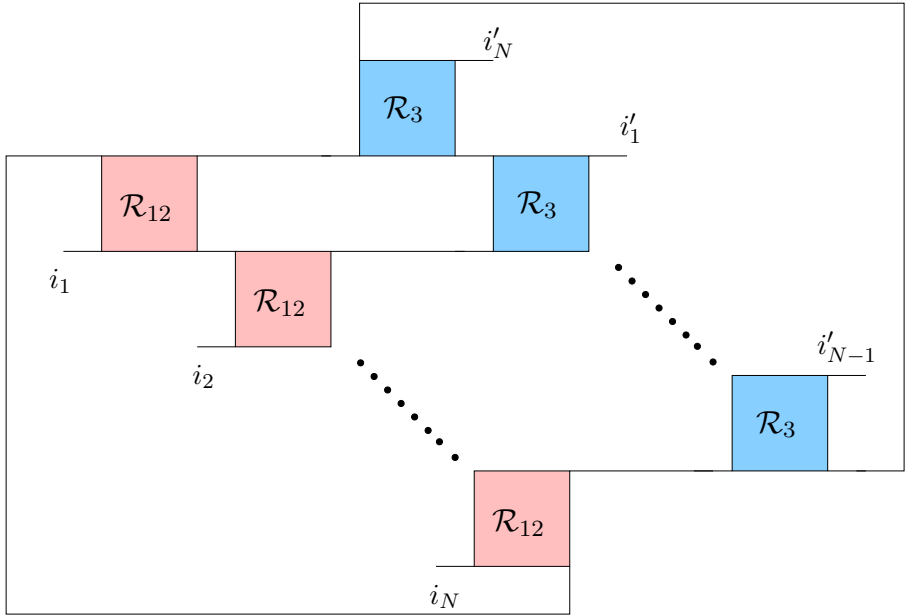}}
\caption[]{The graphical representation of  transfer matrix after transformation.}
\label{fig:fact}
\end{figure}

The  action of the permutation operator
is equivalent to the interchanging of the matrix indices,
$[P_{12} A]^{i_1i_2}_{i'_1i'_2}=A^{i_2i_1}_{i'_1i'_2}$.
Taking this into account one can easily derive the
graphical representation for the transfer matrix~\eqref{TT}, which is  shown in
Fig.~\ref{fig:ladder}.
The boxes with the label $\CR_{12}$ in Fig.~\ref{fig:ladder} denote the kernel of the operator
$\CR_{12}=\CR_1(u-\rho_1+\sigma_1)\CR_2(u-\rho_2+\sigma_2)$, and the boxes with index $\CR_3$
correspond  to the kernel
of the operator $\CR_3(u-\rho_3+\sigma_3)$.
It follows from Eq.~\eqref{RRcomm} that the operators $\CR_{12}$ and $\CR_3$ satisfy the similar
equation
\begin{align}
\CR_{12}^{(ik)}\CR_3^{(kj)}=\CR_3^{(kj)}\CR_{12}^{(ik)}\,.
\end{align}
The graphical representation of the above identity is given by the diagram in
Fig.~\ref{fig:RR} where the box  $\CR_k$ has to be
understood as the matrix for the operator $\CR_{12}$.
Then using this  relation one can bring the diagram in Fig.~\ref{fig:ladder}
into the form shown in Fig.~\ref{fig:fact}.

One can easily check that this diagram is nothing else as
the graphical representation of the following operator
\begin{align}
\left(\tr_0 P_{10}\CR_{12}^{(10)}\ldots P_{N0} \CR_{12}^{(N0)}\right)\mathcal{P}^{-1}
\left(\tr_0 \mathcal{L}^{(3)}_{10}(u_3)\ldots \mathcal{L}^{(3)}_{N0}(u_3)\right)\,,
\end{align}
where $\CR_{12}^{(k0)}=\CR_1^{(k0)}(u-\rho_1+\sigma_1)\CR_2^{(k0)}(u-\rho_1+\sigma_1)$
and $u_3=u-\rho_3+\sigma_3$.
Taking into account Eq.~\eqref{Q123}  one concludes that
last trace corresponds to the operator $Q_3(u+\sigma_3)$.
We get
\begin{align}\label{f-st}
\TT_{\boldsymbol{m}}(u)&=\tr_{0}\left\{P_{10} \left(\CR_1^{(10)}\CR_2^{(10)}\CR_3^{(10)}\right)
\ldots P_{N0} \left(\CR_1^{(N0)}\CR_2^{(N0)}\CR_3^{(N0)}
\right)\right\}=\nonumber\\[3mm]
&\tr_{0}\left\{P_{10} \left(\CR_1^{(10)}\CR_2^{(10)}\right)
\ldots P_{N0} \left(\CR_1^{(N0)}\CR_2^{(N0)}
\right)\right\}\, \mathcal{P}^{-1} Q_3(u+\sigma_3)\,.
\end{align}
The trace in the second line of Eq.~\eqref{f-st} differs from
the  trace  in the first line  by the absence of the operator $\CR_3$ only.
Hence one can repeat the same steps  and show that
\begin{align}\label{2A}
&\tr_{0}\left\{P_{10} \left(\CR_1^{(10)}\CR_2^{(10)}\right)
\ldots P_{N0} \left(\CR_1^{(N0)}\CR_2^{(N0)}
\right)\right\}=\ \ \ \ \ \ \ \ \ \ \ \ \ \ \ \ \ \ \ \ \ \,\nonumber\\[3mm]
&\ \ \ \ \left(\tr_0 \mathcal{L}^{(1)}_{10}(u_1)\ldots \mathcal{L}^{(1)}_{N0}(u_1)\right)
\mathcal{P}^{-1}
\left(\tr_0 \mathcal{L}^{(2)}_{10}(u_2)\ldots \mathcal{L}^{(2)}_{N0}(u_2)\right)=
\mathcal{P}\,Q_1(u+\sigma_1)Q_2(u+\sigma_2) \,.
\end{align}
Thus, one obtains for $\TT_{\boldsymbol{m}}(u)$
$$
\TT_{\boldsymbol{m}}(u)=\mathcal{P}Q_1(u+\sigma_1)\,Q_2(u+\sigma_2)\mathcal{P}^{-1}Q_3(u+\sigma_3)=
Q_1(u+\sigma_1)\,Q_2(u+\sigma_2)Q_3(u+\sigma_3)\,,
$$
where we took into account that $[\mathcal{P},
Q_k(u)]=0$.

Let us note that the construction of the Baxter operators $Q_k(u)$ and the proof of the
factorization of the transfer matrix
presented here rely on the properties of the operators $\CR_k$
only. Therefore, the same constructions will hold for the generic $sl(N)$  spin chain
provided that there exist operators $\CR_k$ satisfying the equations analogous to
Eqs.~\eqref{123LL}.

\section{Transfer matrices for the reducible modules}\label{sect:red-mod}
So far we have  considered the $\CR-$operators which act on the tensor product of two generic
$sl(3)$ modules, $\VV_{\boldsymbol{n}}\otimes\VV_{\boldsymbol{m}}$,
$\VV_{\boldsymbol{n(m)}}\sim \VV=\mathbb{C}[x,y,z]$.
We have shown that the transfer matrix $\TT_{\boldsymbol{m}}(u)$, (see Eq.~\eqref{TT}),
is factorized into the product of three $Q$ operators, provided that the auxiliary space
$\VV_{\boldsymbol{m}}$ is a generic one. In this section we study the properties of the
transfer matrices in the case that the auxiliary space is non-generic. Namely, our final aim
is to find the expression for the transfer matrix with the finite dimensional auxiliary space
in terms of the generic transfer matrix $\TT_{\boldsymbol{m}}(u)$.

The subsequent analysis is based on the generalization of the method used in
Ref.~\cite{DM05} for the analysis of the $sl(2)$ invariant spin chains.   We
briefly remind the main idea and then go over to the more detailed discussion.
First of all we  note that for the proof of factorization
it is
completely irrelevant whether the module $\VV_{\boldsymbol{m}}$ is irreducible or not.
Second,  let us consider the situation when the (auxiliary) module $\VV_{\boldsymbol{m}}$
is reducible, i.e. it contains an invariant subspace, $\widetilde \VV\subset\VV_{\boldsymbol{m}}$.
Then the subspace $\VV_{\boldsymbol{n}}\otimes\widetilde \VV\subset
\VV_{\boldsymbol{n}}\otimes\VV_{\boldsymbol{m}}$  is the invariant subspace of the
 operator $\CR_{\boldsymbol{n}\boldsymbol{m}}(u)$.
As a consequence, the latter has the block
diagonal form
\begin{align}\label{Block}
\CR_{\boldsymbol{n}\boldsymbol{m}}(u)=
\begin{pmatrix}\widetilde\CR(u) &\star\\
0& \widebar\CR(u)
\end{pmatrix}\,.
\end{align}
The new operator $\widetilde\CR(u)$ acts on the space
$\VV_{\boldsymbol{n}}\otimes\widetilde \VV$
and $\widebar\CR(u)$ --- on the space
$\VV_{\boldsymbol{n}}\otimes\widebar  \VV$, where
$\widebar \VV=\VV_{\boldsymbol{m}}/\widetilde \VV $ is the factor space. The operators
 $\widetilde\CR(u)$,  $\widebar\CR(u)$  satisfy the YB relation and can be identified with
the $\CR$ operators on the corresponding spaces. Clearly, the trace in Eq.~\eqref{TT} decays into the
sum of two traces,
\begin{align}\label{TTT}
\TT_{\boldsymbol{m}}(u)=\widetilde\TT(u)+\widebar \TT(u)\,,
\end{align}
where
\begin{align}
\widetilde\TT(u)=\tr\left(\widetilde \CR_{10}(u)\ldots\widetilde \CR_{N0}(u)\right)\,,&&
 \widebar\TT(u)=\tr\left(\widebar \CR_{10}(u)\ldots\widebar \CR_{N0}(u)\right)\,.
\end{align}
If the factor module is isomorphic to a certain
generic module, $\widebar \VV=\VV_{\boldsymbol{m}'}$, one concludes that the transfer matrix
$ \widebar\TT(u)=\varphi(u)\TT_{\boldsymbol{m}'}(u)$, where $\varphi(u)$ is some normalization
factor. Thus, Eq.~\eqref{TTT} allows to express the  new transfer matrix $\widetilde\TT(u)$
in terms of the two generic transfer matrix $\TT_{\boldsymbol{m}}$ and  $\TT_{\boldsymbol{m}'}$.
In what follows we shall show that the all transfer matrices with non-generic auxiliary space can
be expressed as the certain combinations of the generic transfer matrices $\TT_{\boldsymbol{m}}(u)$.
Below we consider the reducible $sl(3)$ modules in more details.

\subsection{Structure of the reducible modules}
\label{subs:modules}
The generic $sl(3)$ module $\VV_{m_1,m_2}$ is irreducible unless one of the numbers
$1-m_1,\, 1-m_2,\, 2-m_1-m_2$ is a positive integer. This can be checked in the following
way. Let us consider the space $\WW=\mathbb{C}[x,y,z]$ and  the bilinear form $(\cdot,\cdot)$
on $\WW\times\VV$, ($\VV\equiv\VV_{m_1,m_2}$) defined as follows
\begin{align}
(\tilde e_{k_1,k_2,k_3},e_{n_1,n_2,n_3})=
\delta_{n_1,k_1}\delta_{n_2,k_2}\delta_{n_3,k_3} n_1!n_2!n_3!
\end{align}
where $e(\tilde e)_{n_1,n_2,n_3}=x^{n_1} y^{n_2} z^{n_3}$ are the basis vectors
in the spaces $\VV(\WW)$. The representation of the $sl(3)$ algebra on $\VV$ induces the
representation on $\WW$, $(w, L_\alpha v)=(\tilde L_\alpha w,v)$. Now if the subspace
$\widetilde \VV$
is an  invariant subspace, $\widetilde \VV\subset\VV$, then the subspace $\WW_\perp$,
orthogonal to $\widetilde \VV$, $(\WW_\perp,\widetilde \VV)=0$, is  the invariant subspace
of $\WW$. Next, it is easy to see that there exists only one
lowest weight vector in the space $\VV$ $\Phi=1$.
Since an invariant subspace has to have a lowest weight vector, $\Phi\in \widetilde \VV$ as well.
Therefore one concludes that the invariant
subspace $\WW_\perp$ has a lowest weight vector $\Psi\neq 1$.
So, if the space  $\VV$ has an invariant subspace there has to exist a nontrivial,
$\Psi\neq 1$, lowest weight vector,
\begin{align}\label{lowest-w}
\tilde L_{ik}\Psi=0,\ \ \ \ \ \  i<k,
\end{align}
in the dual space $\WW$. One can easily find that the equations~\eqref{lowest-w} have a solution
only if at least one  of the numbers  $1-m_1,\, 1-m_2,\, 2-m_1-m_2$ is a positive integer.
If only one of these numbers is a positive integer, then there exists only one
nontrivial solution (lowest weight),
and as a consequence, only one  invariant subspace $\widetilde \VV$ (see
Appendix~\ref{app:modules} for details).

To get a more detail description of the invariant subspaces we define two operators
\begin{align}\label{cd12}
\CD_1=\partial_x+z\partial_y\,,\ \ \ \ \ \ \ \ \CD_2=\partial_z\,.
\end{align}
These operators possess  remarkable properties. Namely, if $1-m_1$ is a positive integer
the operator $\CD_1^{1-m_1}$
intertwines generators $L_\alpha(m_1,m_2)$ and $L_\alpha(2-m_1,m_1+m_2-1)$,
and if  $1-m_2$ is a positive integer
the operator $\CD_2^{1-m_2}$ intertwines generators $L_\alpha(m_1,m_2)$ and
$L_\alpha(m_1+m_2-1,2-m_2)$
\begin{subequations}\label{LL}
\begin{align}
\CD_1^{1-m_1}\,L_\alpha(m_1,m_2)&=L_\alpha(2-m_1,m_1+m_2-1)\,\CD_1^{1-m_1}\,,\\[2mm]
\CD_2^{1-m_2}\,L_\alpha(m_1,m_2)&=L_\alpha(m_2+m_1-1,2-m_2)\,\CD_2^{1-m_2}\,.
\end{align}
\end{subequations}
To check this, note that the  operators $\CD_1$ and $\CD_2$ commute with all lowering generators,
$L_{ik}$, $i>k$. It can be shown that
\begin{align}
\CD_1 L_{12}(m_1)&=L_{12}(m_1+2) \CD_1+m_1\,,&
\CD_1 L_{23}(m_2)&=L_{23}(m_2-1) \CD_1\,,\\[2mm]
\CD_2 L_{12}(m_1)&=L_{12}(m_1-1) \CD_2\,,
&\CD_2 L_{23}(m_2)&=L_{23}(m_2+2) \CD_2+m_2\,.
\end{align}
Here we display only those spins as arguments of the generators  that they really depend
on. From these relations
it follows for example
that
$$
 \CD_1^p  L_{12}(m_1)=L_{12}(m_1+2p) \CD_1^p+p(m_1+p-1)\CD_1^{p-1}
$$
and therefore for
$p=1-m_1$ the inhomogeneous term disappears.
Thus the Eqs.~\eqref{LL} hold for the generators $L_{12}$, $L_{23}$ and the
lowering generators.
Since all other generators can be obtained as the commutators of the latter
ones we conclude that Eqs.~\eqref{LL} are valid for all generators.
It is useful to rewrite Eqs.~\eqref{LL} in $\boldsymbol{\sigma}$ notations.
Taking into account the definitions~\eqref{mk} one gets
\begin{subequations}\label{LLs}
\begin{align}
\CW_{12}\,L_\alpha(\sigma_1,\sigma_2,\sigma_3)&=
L_\alpha(\sigma_2,\sigma_1,\sigma_3)\,\CW_{12}&& \CW_{12}=\CD_1^{\sigma_{12}}\,,\\[2mm]
\CW_{23}\,L_\alpha(\sigma_1,\sigma_2,\sigma_3)&=
L_\alpha(\sigma_1,\sigma_3,\sigma_2)\,\CW_{23}&&\CW_{23}=\CD_2^{\sigma_{23}}\,,
\end{align}
\end{subequations}
where $\sigma_{ik}=\sigma_i-\sigma_k$. Thus the action of these operators
results in the permutation of the "spins" \mbox{$\boldsymbol{\sigma}=(\sigma_1,\sigma_2,\sigma_3)$.}
These relations are valid even for   a non-integer $\sigma_{12}(\sigma_{23})$.
However, the operators $\CD_1^{\sigma_{12}},\CD_2^{\sigma_{23}}$ are well
defined operators on $\VV$ only for  integer $\sigma_{12}, \sigma_{23}$.
Thus, if $\sigma_{23}=1,2,\ldots$ ($m_2=0,-1,-2,\ldots$)  the kernel of the
operator $\CW_{23}=\CD_2^{\sigma_{23}}$
\begin{align}\label{V2}
\VV^{(23)}_{\sigma_1\sigma_2\sigma_3}=\ker\CW_{23}
\end{align}
is an invariant subspace in $\VV_{\sigma_1\sigma_2\sigma_3}$.
The factor space,
$\VV_{\sigma_1\sigma_2\sigma_3}/\VV^{(23)}_{\sigma_1\sigma_2\sigma_3}=
\mathrm{Im}\,\CW_{23}$,
 is the $sl(3)$ module having the "quantum
numbers" $\sigma_1,\sigma_3,\sigma_2$.  For  $\sigma_{12}=1-m_1$
being non-integer this module is irreducible  and, therefore, coincides with the
generic $sl(3)$ module, $\VV_{\sigma_1\sigma_3\sigma_2}$, i.e.
\begin{align}
\VV_{\sigma_1\sigma_2\sigma_3}\xrightarrow{\CD_2^{\sigma_{23}}}\VV_{\sigma_1\sigma_3\sigma_2}=
\VV_{\sigma_1\sigma_2\sigma_3}/\VV^{(23)}_{\sigma_1\sigma_2\sigma_3}\,.
\end{align}
 Since the
operator $\CD_2^{\sigma_{23}}$ does not depends on $m_1$, this conclusion remains valid for
an arbitrary $m_1$.
Thus, for $-m_2=0,1,2,\ldots$ the generic module
$\VV_{\sigma_1\sigma_2\sigma_3}$ contains the invariant subspace~\eqref{V2},
and the factor space $\VV/\VV^{(23)}$ is the generic module
$\VV_{\sigma_1\sigma_3\sigma_2}$. For non-integer $\sigma_{12}=1-m_1$
 these modules are irreducible ones.

Similarly, if $\sigma_{12}=1-m_1=1,2,\ldots$ the generic module
$\VV_{\sigma_1\sigma_3\sigma_2}$ contains the invariant subspace,
$\VV^{(12)}_{\sigma_1\sigma_2\sigma_3}=\ker{\CW_{12}}$,
and the
factor space $\VV/\VV^{(12)}=\mathrm{Im}\CW_{12}$ is equivalent to
the generic module $\VV_{\sigma_2\sigma_1\sigma_3}$,
\begin{align}
\VV_{\sigma_1\sigma_2\sigma_3}\xrightarrow{\CD_1^{\sigma_{12}}}\VV_{\sigma_2\sigma_1\sigma_3}=
\VV_{\sigma_1\sigma_2\sigma_3}/\VV^{(12)}_{\sigma_1\sigma_2\sigma_3}\,.
\end{align}
Again,  these modules are
irreducible ones provided that  $\sigma_{23}$ is non-integer.

Now let us consider the situation when $\sigma_{13}=2-m_1-m_2$ is a positive
integer, while $\sigma_{12}=1-m_1$ and $\sigma_{23}=1-m_2$ are not. In this case there
exists operator $\mathcal{W}$ which intertwines the generators
$L_\alpha(\sigma_1\sigma_2\sigma_3)$ and
$L_\alpha(\sigma_3\sigma_2\sigma_1)$. It has the form
\begin{align}\label{W}
\CW_{13}=\CD_2^{\sigma_{12}}\CD_1^{\sigma_{13}}\CD_2^{\sigma_{23}}=
\CD_1^{\sigma_{23}}\CD_2^{\sigma_{13}}\CD_1^{\sigma_{12}}\,.
\end{align}
Indeed, using relations~\eqref{LLs} it is easy to check that the operator
$\CW_{13}$ (in both forms) intertwines the corresponding operators. Next, taking into
account that the commutator $[\CD_1,\CD_2]$ commutes with both $\CD_1$ and $\CD_2$
it is straightforward to show that $\CW_{13}$  is a polynomial in $\CD_1$
and $\CD_2$, and that both representations are equivalent. One
concludes that the kernel of the operator $\CW_{13}$ is an invariant
subspace, $\VV_{\boldsymbol{\sigma}}^{(13)}=\ker \CW_{13}$, and the
corresponding factor module is equivalent to the generic module
$\VV_{\sigma_3,\sigma_2,\sigma_1}$,
\begin{align}
\VV_{\sigma_1\sigma_2\sigma_3}\xrightarrow{\CW_{13}}\VV_{\sigma_3\sigma_2\sigma_1}=
\VV_{\sigma_1\sigma_2\sigma_3}/\VV^{(13)}_{\sigma_1\sigma_2\sigma_3}\,.
\end{align}

If one of the numbers
$\sigma_{12},\sigma_{23},\sigma_{13}$ is positive integer
we define the transfer matrices $\TB^{(ij)}_{\bs}$, $i<j$ by
\begin{align}
\TB^{(ij)}_{\bs}(u)=\tr_{\VV^{(ij)}_{\bs}}\left\{\widetilde\CR_{10}(u)
\ldots\widetilde\CR_{N0}(u)\right\}\,,
\end{align}
where $\widetilde\CR_{k0}(u)$ is the restriction of the operator $\CR_{k0}(u)$
on the subspace $\VV_k\otimes \VV^{(ij)}_{\bs}$. As it was explained in the beginning of
the section, the second diagonal block of the $\CR-$operator, Eq.~\eqref{Block}, is
proportional to the $\CR$ operator on the tensor product
$\VV_{\br}\otimes\left(\VV_{\bs}/\VV^{(ij)}_{\bs}\right)$.
Since $\VV_{\bs}/\VV^{(ij)}_{\bs}=\VV_{\bs_{ij}}$, where $\bs_{ij}=P_{ij}\bs$
(the operator $P_{ij}$
interchanges the $i-$th and $j-$th components of the vector $(\sigma_1,\sigma_2,\sigma_3)$)
one concludes that
\begin{align}\label{twineR}
\CW_{ij}\,\widebar\CR(u)~=~r_{ij}(u)\,
\CR_{\br\,\bs_{ij}}(u)\,\CW_{ij}.
\end{align}
To fix the factor $r_{ij}(u)$ it is sufficient to apply the l.h.s. and r.h.s. to some vector.
With the help of the formulae given in Appendix~\ref{app:operators} it can
be checked that for the chosen
normalization of the $\CR-$operator all  coefficients $r_{ij}(u)$ are equal to $1$.
Then taking into account \eqref{TTT} one obtains
\begin{align}\label{1}
\TB^{(ij)}_{\bs}(u)=\TT_{\bs}(u)-\TT_{\bs_{ij}}(u)=(1-P_{ij})\TT_{\bs}(u)\,.
\end{align}
\vskip 0.5cm

In the case that both spins $m_1,m_2$ are negative, $m_1,m_2=0,-1,-2\ldots$, or,
equivalently, the differences $\sigma_{12},\sigma_{23}$ are positive
integers, the subspaces $\VV^{(ij)}_{\bs}$ are not irreducible any longer.
Thus one can again single out an invariant subspace and write for the transfer matrix
$ \TB^{(ij)}_{\bs}(u)$ the representation similar to  Eq.~\eqref{TTT}.
For  definiteness we consider the space $\VV^{(23)}_{\bs}$.
As follows from Eqs.~\eqref{cd12} and \eqref{LLs} this is the vector space spanned
by the  basis vectors $x^nz^ky^p$,
such that $k\leq -m_2=\sigma_{23}-1$. Using the arguments given at the
beginning of the subsection (see the discussion around Eq.~\eqref{lowest-w} ) one can show
that this space contains only one invariant subspace (see Appendix~\ref{app:modules} for details).
This invariant subspace, $v_{\bs}$,
is given by the kernel of the operator $\CW_{12}$ restricted on   $\VV^{(23)}_{\bs}$,
or by the intersections of the kernels of the operators $\CW_{12}$ and $\CW_{23}$,
\begin{align}
v_{\bs}=\ker\CW_{12}\bigl|_{\VV^{(23)}_{\bs}}= \ker\CW_{12}\cap\ker\CW_{23}=
\ker\CW_{23}\bigl|_{\VV^{(12)}_{\bs}}\,.
\end{align}
Obviously, the module $v_{\bs}$ is the finite dimensional $sl(3)$ module.
As was discussed in the beginning of the section, the $\CR$ matrix
takes the block diagonal form~\eqref{Block},
where $\widetilde\CR(u)$ is now the restriction of the $\CR_{\br\bs}(u)$ operator on the
invariant subspace $\VV_{\br}\otimes v_{\bs}$, and $ \widebar\CR(u) $ is the
restriction of the operator $\CR_{\br\bs}(u)$ on the subspace
$\VV_{\br}\otimes\left(\VV^{(23)}_{\bs}/v_{\bs}\right)$,
\begin{align}
\VV^{(23)}_{\bs}/v_{\bs}\sim
\mathrm{Im}\CW_{12}\bigl|_{\VV^{(23)}_{\bs}}\equiv  \CV_{\sigma_2\sigma_1\sigma_3}.
\end{align}
As a consequence, one obtains the following relation for the transfer matrices
\begin{align}\label{2}
\TB^{(23)}_{\sigma_1\sigma_2\sigma_3}(u)=t_{\sigma_1\sigma_2\sigma_3}(u)+
\CT_{\sigma_2\sigma_1\sigma_3}(u)\,.
\end{align}
The new transfer matrices entering the Eq.~\eqref{2} are defined as follows
\begin{align}
t_{\bs}(u)&=\tr_{v_{\bs}}\left\{\widetilde\CR_{10}(u)\ldots\widetilde \CR_{N0}(u)
\right\}\,,\\[2mm]
\CT_{\bs_{12}}(u)&=\tr_{\CV_{{\bs}_{12}}}
\left\{\widebar \CR_{10}(u)\ldots \widebar\CR_{N0}(u)\right\}\,.
\end{align}

At the last step we  express the transfer matrix
$\CT_{\bs}$ in terms of $\TB^{(23)}_{\bs}$. To this end we note that
the  module
$\CV_{\sigma_2\sigma_1\sigma_3}=
\mathrm{Im}\CW_{12}\bigl|_{\VV^{(23)}_{\bs}} $ is contained in the module
$\VV^{(23)}_{\sigma_2\sigma_1\sigma_3}$,
$\CV_{\sigma_2\sigma_1\sigma_3}\subset\VV^{(23)}_{\sigma_2\sigma_1\sigma_3}$.
To verify this one
should show that $\CD_2^{\sigma_{13}} \varphi=0$ if $\varphi\in\CV_{\sigma_2\sigma_1\sigma_3}$.
Indeed, taking into account that $\varphi=\CD_1^{\sigma_{12}} f$, where
$f\in\VV^{(23)}_{\sigma_1\sigma_2\sigma_3}$, ( $D_{2}^{\sigma_{23}}f=0$ ) and that
$[\CD_2[\CD_1,\CD_2]]=0$
one gets
\begin{align}
\CD_2^{\sigma_{13}}\varphi=
\CD_2^{\sigma_{13}}\CD_1^{\sigma_{12}} f=\widetilde W\, \CD_2^{\sigma_{23}}f=0\,.
\end{align}
Next, this subspace, $\CV_{\sigma_2\sigma_1\sigma_3}$ coincides with the kernel of the
intertwining operator~\footnote{Let us note that though $\sigma_{21}<0$
the operator $\widetilde\CW_{13}$ is a polynomial in
$\CD_1,\CD_2$.}
\begin{align}\label{W13}
\widetilde\CW_{13}=\CD_2^{\sigma_{21}}\CD_1^{\sigma_{23}}\CD_{2}^{\sigma_{13}}=
\CD_1^{\sigma_{13}}\CD_2^{\sigma_{23}}\CD_1^{\sigma_{21}}\,,&&
\widetilde\CW_{13}L_\alpha(\sigma_2,\sigma_1,\sigma_3)=
L_\alpha(\sigma_3,\sigma_1,\sigma_2)\widetilde\CW_{13}\,,
\end{align}
i.e. $\CV_{\sigma_2\sigma_1\sigma_3}=\ker\widetilde\CW_{13} $.
Indeed,   it can be shown (see Appendix~\ref{app:modules}) that
the space $\VV^{(23)}_{\sigma_2\sigma_1\sigma_3}$ has only one invariant
subspace. Since both   $\CV_{\sigma_2\sigma_1\sigma_3}$ and
$\ker\widetilde\CW_{13}$ are the invariant subspaces
of $\VV^{(23)}_{\sigma_2\sigma_1\sigma_3}$ they have to coincide.
The factor module
\begin{align}
\VV^{(23)}_{\sigma_2\sigma_1\sigma_3}/\CV_{\sigma_2\sigma_1\sigma_3}=
\VV^{(23)}_{\sigma_2\sigma_1\sigma_3}/\ker\widetilde\CW_{13}=\mathrm{Im}\widetilde\CW_{13}=
\VV^{(23)}_{\sigma_3\sigma_1\sigma_2}
\end{align}
is an irreducible one because neither $\sigma_{31}$ nor $\sigma_{32}$ are not positive integers.
These results is equivalent to the statement that the following sequence
\begin{align}
0{\longrightarrow}v_{\sigma_1\sigma_2\sigma_3}\overset{d_1}{\longrightarrow}
\VV^{(23)}_{\sigma_1\sigma_2\sigma_3}   \overset{
d_2}{\longrightarrow}
\VV^{(23)}_{\sigma_2\sigma_1\sigma_3}   \overset{
d_3}{\longrightarrow}
\VV^{(23)}_{\sigma_3\sigma_1\sigma_2}{\longrightarrow}0\,,
\end{align}
where $d_1=i$ is the natural inclusion of $v_{\bs}$ to
$\VV^{(23)}_{\bs}$, $d_2=\CW_{12}$ and $d_3=\widetilde\CW_{13}$,
is an exact one.
The map $d_2$ in this sequence results in the relation~\eqref{2} for the transfer
matrices,
while the map $d_3$ generates the new relation
\begin{align}   \label{3}
\TB^{(23)}_{\sigma_2\sigma_1\sigma_3}(u)=
\CT_{\sigma_2\sigma_1\sigma_3}(u)+\TB^{(23)}_{\sigma_3\sigma_1\sigma_2}(u)\,.
\end{align}
This equation together  with Eq.~\eqref{2} gives
\begin{align}
t_{\sigma_1\sigma_2\sigma_3}(u)=\TB^{(23)}_{\sigma_1\sigma_2\sigma_3}(u)-
\TB^{(23)}_{\sigma_2\sigma_1\sigma_3}(u)+\TB^{(23)}_{\sigma_3\sigma_1\sigma_2}(u)=
\Bigl(1-P_{12}+P_{12}P_{23}\Bigr)\,\TB^{(23)}_{\sigma_1\sigma_2\sigma_3}(u)\,.
\end{align}
Finally, taking into account Eq.~\eqref{1} we obtain the following representation for
the  auxiliary transfer matrix
\begin{align}\label{t-det}
t_{\sigma_1\sigma_2\sigma_3}(u)=\sum_{P} (-1)^{\mathrm{sign}(P)}\,\,
\TT_{\sigma_{i_1}\sigma_{i_2}\sigma_{i_3}}(u)\,,
\end{align}
where sum is taken over all permutations. We remind that
$\sigma_1>\sigma_2>\sigma_3$ and the differences
$\sigma_{12}=\sigma_1-\sigma_2$, $\sigma_{23}=\sigma_{2}-\sigma_3$ are positive integers.

\section{Baxter equations}\label{sect:baxter}
In this section we shall show that the operators $Q_i(u)$ which factorize the
transfer matrices, $\TT_{\bs}(u)$, satisfy the infinite set of  finite difference
relations involving the auxiliary (finite dimensional) transfer matrices
$t_{\bs}(u)$.

First of all let us note that using the factorized form of the transfer matrix,
Eq.~\eqref{T3Q}, one can rewrite Eq.~\eqref{t-det} in the equivalent form
\begin{align}\label{tq-det}
t_{\sigma_1\sigma_2\sigma_3}(u)=\det||Q_i(u+\sigma_j)||_{j,i=1,2,3}\,.
\end{align}
All linear equations on the operators $Q_k$ can be obtained in the following way.
Let us consider the determinant of the matrix which has two identical columns
\begin{align}\label{qik}
\det\left|
\begin{array}{cccc}
  {Q}_1(u+\sigma_1) & {Q}_2(u+\sigma_1)
  & {Q}_3(u+\sigma_1)& {Q}_3(u+\sigma_1)\\
  {Q}_1(u+\sigma_2) & {Q}_2(u+\sigma_2)
  & {Q}_3(u+\sigma_2)& {Q}_3(u+\sigma_2)\\
  {Q}_1(u+\sigma_3) & {Q}_2(u+\sigma_3)
  & {Q}_3(u+\sigma_3)& {Q}_3(u+\sigma_3)\\
  {Q}_1(u+\sigma_4) & {Q}_2(u+\sigma_4)
  & {Q}_3(u+\sigma_4)& {Q}_3(u+\sigma_4)
\end{array}
\right|=0\,.
\end{align}
We assume that the parameter $\sigma_4$ is such that the difference
$\sigma_{34}=\sigma_3-\sigma_4\equiv 1-m_3$
is an positive integer.
Expanding the determinant \eqref{qik} over the last column one derives
\begin{align}\label{Bax}
\sum_{k=1}^{4}(-1)^k\,Q_3(u+\sigma_k) \det{M_k}(u)=0\,,
\end{align}
where $M_k(u)$ are the corresponding minors. The minors are the $3\times3$
matrices of the same type as in \eqref{tq-det} and can be identified with the
transfer matrices as follows
\begin{align}\label{M-t}
M_k(u)=t_{\bs_k(\alpha)}
\left(u+\alpha\right)|_{\alpha=(\sigma_4-\sigma_k)/3}\,,
\end{align}
where $\bs_k(\alpha)=(\sigma_1-\alpha,..\hat \sigma_k,..\sigma_4-\alpha)$,
i.e. all parameters $\sigma_i$, $i=1,\ldots,4$ are shifted by $\alpha$, and the $k-th$ element
in the string $(\sigma_1,\sigma_2,\sigma_3,\sigma_4)$ is omitted. The shift
of the arguments arises from the requirement that
$\sum_{i=1}^3\sigma^k_i(\alpha)=0$. At this point it is convenient to return to the
standard notations for the transfer matrix, $t_{\bs}(u)=t_{\boldsymbol{m}}$,
$m_k=\sigma_{k+1}-\sigma_{k}+1$.  Then the equation \eqref{Bax} takes form
\begin{align}
&t_{m_2,m_3}\left(u+\frac{m_1+m_2+m_3-3}{3}\right) Q_3(u+\sigma_1)-
t_{m_1+m_2-1,m_3}\left(u+\frac{m_2+m_3-2}{3}\right) Q_3(u+\sigma_2)+\nonumber\\
&t_{m_1,m_2+m_3-1}\left(u+\frac{m_3-1}{3}\right)Q_3(u+\sigma_3)
-t_{m_1,m_2}(u) Q_3(u+\sigma_4)=0\,,
\end{align}
where the parameters $\sigma_k$ are given by
\begin{align}
\sigma_1=1-\frac{2m_1+m_2}{3},&&\sigma_2=\frac{m_1-m_2}{3},&&\sigma_3=-1+\frac{m_1+2m_2}{3},&&
\sigma_4=-2+\frac{m_1+2m_2+3m_3}{3}\,.
\end{align}
The parameters $m_i$ take the following values: $m_i=0,-1,-2,\ldots$.
Obviously, the  other two operators $Q_2(u)$ and $Q_3(u)$ satisfy the same equation.
Having put all $m_i=0$ one  gets the simplest equation on the operator $Q_k(u)$
\begin{align}   \label{BaxterEq}
&t_{(0,0)}\left(u-1\right) Q_k(u+1)+t_{(0,-1)}\left(u-\frac{1}{3}\right)Q_k(u-1)=
t_{(-1,0)}\left(u-\frac{2}{3}\right) Q_k(u)+
t_{(0,0)}(u) Q_k(u-2)\,.
\end{align}
The auxiliary transfer matrices $t_{0,-1}$ and $t_{-1,0}$ can be represented as the traces
of the Lax operators
\begin{align}
L(u)=L_{(0,-1)}(u)=u+\sum_{ab} E_{ba} L_{ab}\,,&&
\widebar L(u)=L_{(-1,0)}(u)=u+\sum_{ab} \widebar E_{ba} L_{ab}\,,
\end{align}
where $E_{ab}$ are the generators in the fundamental representation of $sl(3)$,
($\boldsymbol{m}=(0,-1)$), and $\widebar E_{ba}$ -- are the generators in the
representation $\boldsymbol{m}=(-1,0)$. Let us denote by
$\widetilde \CR_{\boldsymbol{n}\boldsymbol{m}}$
the restriction of the $\CR$ operators to the invariant subspaces for
$\boldsymbol{m}=(0,-1)$ or $\boldsymbol{m}=(-1,0)$.
Using the formulae from  Appendix~\ref{app:operators} one finds by comparison of the
eigenvalues that
\begin{align}
\widetilde \CR_{\br,(0,-1)}(u-1/3)=-X(u)L(u)\,,&&
\widetilde \CR_{\br,(-1,0)}(u-2/3)=X(u){\bar L}^{-1}(-u+1)\,,
\end{align}
where
\begin{align}
X(u)=\cos\pi(u-\rho_2)\frac{\Gamma(u+1-\rho_1)\Gamma(u-1-\rho_1)\Gamma(u-1-\rho_2)}{
\Gamma(\rho_2-u)\Gamma(\rho_3-u)\Gamma(\rho_3-u+2)}\,.
\end{align}
Similarly, one derives
\begin{align}
\widetilde \CR_{\br,(0,0)}(u)= X(u)\,\prod_{k=0}^3(\rho_k+1-u)\,.
\end{align}
Then Eq.~\eqref{BaxterEq} can be rewritten in the form
\begin{align}\label{Bk}
\tau_2(u)\,Q_k(u)+\Delta(u)\,\Delta(u-1)\,Q_k(u-2)=\Delta(u)\, \tau_1(u)\, Q_k(u-1)+Q_k(u+1)\,,
\end{align}
where
$\Delta(u)=\prod_{k=1}^3(u-\rho_k)^N$
and the the auxiliary transfer matrices $\tau_1(u)$ and $\tau_2(u)$ are the polynomials in $u$ of
degree $N$ and $2N$
\begin{align}
\tau_1(u)=&\tr\left\{L_1(u)\ldots L_N(u)\right\}=3u^{N}+\sum_{k=2}^{N} q_{k}^{(1)}\,u^{N-k}
\,,\\[2mm]
\tau_2(u)=&\prod_{k=1}^3(\rho_k-u)^N\tr\left\{\bar L_1^{-1}(-u+1)\ldots \bar
L_N^{-1}(-u+1)\right\}=
3u^{2N}+3N\,u^{2N-1}+\sum_{k=2}^{2N} q_{k}^{(2)}\,u^{2N-k}
\end{align}
The charges $q_k^{(i)}$ are some functions of the spin generators. The lowest charges
can be expressed in terms of the  Casimir operators as follows
\begin{subequations}
\begin{align}
q_2^{(1)} =&\mathbb{C}_2^N-N\mathbb{C}_2\,,\\[2mm]
q_2^{(2)}=&\mathbb{C}_2^N-2N\mathbb{C}_2
+\frac32N(N-1)\,,\\[2mm]
q_3^{(2)}=&q_3^{(1)}-[\mathbb{C}_3^N-N\mathbb{C}_3]+N(\mathbb{C}_2^N-N\mathbb{C}_2)-N(N-1)\mathbb{C}_2
+\frac12N(N-1)(N-2)
\end{align}
\end{subequations}
Here the operators $\mathbb{C}_2 (\mathbb{C}_3)$ and $\mathbb{C}_2^N (\mathbb{C}_3^N)$
are the ``one-particle'' and the total quadratic (cubic) Casimir operators,
\begin{align}
\mathbb{C}_2=\frac12 L_{ab}L_{ba},&&
\mathbb{C}_2^N=\frac12 \mathbf{L}_{ab}\mathbf{L}_{ba}\,,&&
C_3=\frac13 L_{ab}L_{bc}L_{ca} \,,&& C_3^N=\frac13 \mathbf{L}_{ab}\mathbf{L}_{bc}\mathbf{L}_{ca}\,,
\end{align}
where $\mathbf{L}_{ab}=(L_1+\ldots+L_N)_{ab}$.

Further, let us define  new operator $Q(u)$ as
\begin{align}\label{QQ1}
Q_3(u)=\Gamma^N(u-\rho_2+1)\Gamma^N(u-\rho_1+1) Q(u)\,.
\end{align}
It can be shown that the eigenvalues of the operator $Q(u)$ are  polynomials in $u$.
Inserting the ansatz ~\eqref{QQ1} into Eq.~\eqref{Bk} one derives
\begin{align}\label{beq}
&\tau_2(u)\, Q(u)+(u-\rho_3)^N(u-\rho_3-1)^N \,Q(u-2)=\nonumber\\[2mm]
&(u-\rho_3)^N \tau_1(u) \,Q(u-1)+ (u-\rho_1+1)^N(u-\rho_2+1)^N \, Q(u+1)\,.
\end{align}
The degree of the polynomial $Q(u)$ is determined by the eigenvalues of the Cartan generators
$\mathbf{H}_1$ and $\mathbf{H}_2$. Namely, $Q(u)\sim u^M+\ldots$, where
$
M=\frac13\left(2H_2+H_1-N(2n_2+n_1)\right)\,.
$
Clearly, this equation is insufficient to fix the eigenvalues of
all integrals of motion, $q_k^{(i)}$.

To get another equation and establish the connection with Nested Bethe Ansatz let us consider
the operator $Q_2(u)$. Again, separating the ``kinematical'' factor
\begin{align}\label{QQ2}
Q_2(u)=\left(\cos(\pi(u-\rho_2))\frac{\Gamma(u-\rho_1+1)}{\Gamma(1-u+\rho_3)} \right)^N\widehat Q(u)
\end{align}
and taking into account the explicit expression for the operator $\CR_2$, Eq.~\eqref{R2},
one finds that the eigenvalues $\widehat Q(u)$ are  meromorphic functions of $u$ with poles
at the points $u_k=\rho_3+k$, $k=1,2.$. The operator $ \widehat Q(u) $ satisfies a finite
difference equation similar Eq.~\eqref{beq}. Solving  this equation (together with the Eq.~\eqref{beq})
in the class of  meromorphic functions described above one can, in principle,
fix the eigenvalues of all integral of motions.

However, it is more instructive to consider the following operator
\begin{align}   \label{QQ23}
Q_{23}(u)=Q_{3}(u)Q_2(u-1)-Q_{3}(u-1)Q_2(u)\,.
\end{align}
Using  Eq.~\eqref{Bk} one obtains that the operator $Q_{23}(u)$ satisfies the following equation
\begin{align}\label{QQ23E}
\tau_2(u-1)\,Q_{23}(u-1)+&\Delta^{-1}(u)\Delta^{-1}(u-1)\,Q_{23}(u+1)=\nonumber\\[2mm]
&\Delta^{-1}(u-1)\, \tau_1(u)\, Q_{23}(u)+\Delta(u-1)\Delta(u-2)\, Q_{23}(u-2)\,.
\end{align}
Substituting Eqs.~\eqref{QQ1} and~\eqref{QQ2}  into \eqref{QQ23} one gets
\begin{align}\label{qtilde}
Q_{23}(u)=\left(\cos(\pi(u-\rho_2))
\frac{\Gamma(u-\rho_1+1)\Gamma(u-\rho)\Gamma(u-\rho_2)}{\Gamma(1-u+\rho_3)}\right)^N
\widetilde Q(u)\,,
\end{align}
where
\begin{align}
\widetilde Q(u)=Q(u)\widehat Q(u-1)\left(\frac{u-\rho_1}{u-\rho_3-1}\right)^N-Q(u-1)\widehat
Q(u)\,.
\end{align}
Inserting  the ansatz~\eqref{qtilde} into Eq.~\eqref{QQ23E} one gets the following
equation on $\widetilde Q(u)$
\begin{align}\label{secB}
\tau_2(u-1)\,\widetilde Q(u-1)+&(u-\rho_1)^N(u-\rho_1+1)^N \,\widetilde Q(u+1)=\nonumber\\[2mm]
&(u-\rho_1)^N\, \tau_1(u)\, \widetilde Q(u)+(u-\rho_2-1)^N(u-\rho_3-1)^N\, \widetilde Q(u-2)\,.
\end{align}
Let us show that the function $\widetilde Q(u)$ is a polynomial in $u$. Indeed, as we have
shown the function $\widehat Q(u)$, and as a consequence,  $\widetilde Q(u)$
is a meromorphic function with poles at the points $u_k^+=\rho_3+k$, $k=1,2,\ldots$.
Thus, the function $\widetilde Q(u)$
is an analytic function for $\mathrm{Re}(u)<\mathrm{Re}(\rho_3+1)$. Further, it follows from
Eq.~\eqref{secB} that $\widetilde Q(u)$ is a meromorphic function  with poles at the points
$u_k^-=\rho_1+k$, $k=1,2,\ldots$.
For  $\rho_3\neq\rho_1$ we arrive  to the conclusion that the function $\widetilde
Q(u)$ has no poles at all and, therefore,   is  a polynomial.
It can be found from Eq.~\eqref{secB}
that the degree of the polynomial $\widetilde Q(u)$,
(~$\widetilde
Q(u)\sim u^{\widetilde M}+\ldots$~)
  is equal to
$\widetilde M=\frac13(2H_1+H_2-N(2n_1+n_2))$.

In the case of the $sl(3)$ spin magnet
the Nested Bethe Ansatz equations~\cite{KR82,KR83} can be rewritten in the form of the finite
difference equation for two polynomials~\cite{Sklyanin-sl3}. It is easy to check
that Eqs.~\eqref{beq},\eqref{secB} concide with the equations obtained in~\cite{Sklyanin-sl3}.
Let us note that the determinant representation for the auxiliary transfer matrices $\tau_{1,2}(u)$
for the compact $sl(3)$ magnet with the quantum space $(\otimes\mathbb{C}^{\,3})^N$
was derived in~\cite{PS}.

\section{Summary}\label{sect:summary}
We have considered the problem of constructing of the Baxter operators for
the  $sl(3)$ invariant spin magnet.
It has been  shown that the transfer
matrices for the spin
chain with a generic quantum space are factorized into the product of three Baxter
operators.
These operators can be identified with the generic transfer matrix with the special
auxiliary space. We have shown that the transfer matrices with nongeneric
auxiliary space can be represented as  the linear combinations of the generic transfer matrices.
The form of such a representation
is uniquely fixed by the decomposition of the reducible $sl(3)$ modules.
It has been shown that the Baxter operators satisfy the infinite set of the self-consistency
relations (finite-difference equations)
involving the transfer matrices with a finite dimensional auxiliary space;
these equations can be cast into the form equivalent to the Nested Bethe
Ansatz.
Since the approach presented here  does not depend on the existence of the lowest weight
vector in the quantum space of the model, it can be  applied to the analysis of the spin
magnets  of another type, e.g.  the spin magnets
with Hilbert space being the principal series representation of the $sl(n,C)$
group.

\section*{Acknowledgment}
The authors are grateful to G.~P.~Korchemsky for helpful discussions.
This work was supported by
the RFFI grant 05-01-0092 (S.~D.)  and by the Helmholtz Association,
contract number VH-NG-004 (A.~M.).

\appendix

\section{Appendix: $\CR-$operators}
\label{app:operators}
Here we give the explicit expressions for the operators $\CR_k$,
Eq.~\eqref{CRK},
\begin{align}\label{R1}
\CR_1=&\mathbb{S}_1^{-1}\frac{\Gamma(x\partial_x+u_1-v_2+1)}{\Gamma(x\partial_x+1)}
e^{\frac{y}{x}{\partial_z}}\frac{\Gamma(y\partial_y+u_1-v_3+1)}{\Gamma(y\partial_y+v_1-v_3+1)}
 e^{-\frac{y}{x}{\partial_z}}\frac{\Gamma(x\partial_x+1)}{\Gamma(x\partial_x+v_1-v_2+1)}\mathbb{S}_1\,,
 \\[3mm]
\label{R2}
\CR_2=&f(u_2-v_2)\,\,\mathbb{S}_2^{-1}\,
\frac{\Gamma(z_2\partial_{z_2}+u_2-v_3+1)}{\Gamma(z_2\partial_{z_2}+1)}\,
e^{-\frac{y_1}{z_2}{\partial_{x_1}}}\nonumber\\[2mm]
&\ \ \ \ \frac{\Gamma(x_1\partial_{x_1}+u_1-v_2+1)}{\Gamma(x_1\partial_{x_1}+u_1-u_2+1)}
 e^{\frac{y_1}{z_2}{\partial_{x_1}}}\frac{\Gamma(z_2\partial_{z_2}+1)}{\Gamma(z_2\partial_{z_2}
 +v_2-v_3+1)}\mathbb{S}_2\,,\\[3mm]
\CR_3=&\mathbb{S}_3^{-1}\frac{\Gamma(z_1\partial_{z_1}+u_2-v_3+1)}{\Gamma(z_1\partial_{z_1}+1)}
e^{-\frac{y_1}{z_1}{\partial_{x_1}}}
\frac{\Gamma(y_1\partial_{y_1}+u_1-v_3+1)}{\Gamma(y_1\partial_{y_1}+u_1-u_3+1)}
 e^{\frac{y_1}{z_1}{\partial_{x_1}}}
 \frac{\Gamma(z_1\partial_{z_1}+1)}{\Gamma(z_1\partial_{z_1}+u_2-u_3+1)}\mathbb{S}_3\,.
\end{align}
Here $f(\lambda)=\cos\pi\lambda$,
\begin{align*}
x=x_2,&&z=z_2,&& y=y_2-z_2x_2,&&
\partial_x=\partial_{x_2}+z_{2}\partial_{y_2},&&\partial_z=\partial_{z_2}+x_{2}\partial_{y_2},&&
\partial_y=\partial_{y_2}
\end{align*}
and
\begin{align}\label{S-op}
\mathbb{S}_1&=\exp\{(y_1+z_1x_2)\partial_{y_2}\}\,
\exp\{z_1\partial_{z_2}\}\,\exp\{x_1\partial_{x_2}\}\,,\\[2mm]
\mathbb{S}_2&=\exp\{(y_2+z_1x_1)\partial_{y_1}\}\,
\exp\{z_1\partial_{z_2}\}\,\exp\{x_2\partial_{x_1}\}\,,\\[2mm]
\mathbb{S}_3&=\exp\{(y_2+z_2x_1)\partial_{y_1}\}\,
\exp\{z_2\partial_{z_1}\}\,\exp\{x_2\partial_{x_1}\}\,
\end{align}
We remind also that $u_k=u-1-\rho_k$ and $v_k=v-1-\sigma_k$.
The detailed discussion of the properties of these operators can be found in
Ref.~\cite{SD1}.
Since the operators  $\CR_k$ are  $sl(3)$ covariant operators they map
lowest weight vectors to the lowest weight vectors. The latter have the form
\begin{align}\label{lwv}
\Psi_{nmp}=(x_1-x_2)^{n}(z_1-z_2)^{m}(y_1-y_2-z_2(x_1-x_2))^{p}\,.
\end{align}
The Cartan generators have the following values on  these vectors
\begin{align}\label{HPSI}
H_1\Psi_{nmp}=(2n+p-m+n_1+m_1)\Psi_{nmp}\,, &&
H_2\Psi_{nmp}=(2m+p-n+n_2+m_2)\Psi_{nmp}\,.
\end{align}
The vectors $\Psi_{n00}$ and $\Psi_{0m0}$  have  unique quantum numbers and
therefore are the eigenvectors of all operators $\CR_k$.
Denoting the eigenvalues of the operators $\CR_k$ on the vectors $\Psi_{n00}$ and $\Psi_{0m0}$
by $r_k^{(1,n)}$ and $r_k^{(2,m)}$ one gets for the latter
\begin{align}\label{rr}
r_1^{(1,n)}(u_1|v_1,v_2,v_3)&=\frac{\Gamma(n+u_1-v_2+1)}{\Gamma(n+v_1-v_2+1))}
\frac{\Gamma(u_1-v_3+1)}{\Gamma(v_1-v_3+1)}\,,
&r_1^{(2,m)}&=r_1^{(1,0)}\,,\\[2mm]
r_3^{(2,m)}(u_1,u_2,u_3|v_3)&=\frac{\Gamma(m+u_2-v_3+1)}{\Gamma(m+u_2-u_3+1)}
\frac{\Gamma(u_1-v_3+1)}{\Gamma(u_1-u_3+1)}
\,,
&r_3^{(1,n)}&=r_3^{(2,0)}\,,\\[2mm]
r_2^{(1,n)}(u_1,u_2|v_2,v_3)&=f(u_2-v_2)\,\frac{\Gamma(n+u_1-v_2+1)}{\Gamma(n+u_1-u_2+1)}
\frac{\Gamma(u_2-v_3+1)}{\Gamma(v_2-v_3+1)}\,,\\[2mm]
r_2^{(2,m)}(u_1,u_2|v_2,v_3)&=f(u_2-v_2)\,\frac{\Gamma(u_1-v_2+1)}{\Gamma(u_1-u_2+1)}
\frac{\Gamma(m+u_2-v_3+1)}{\Gamma(m+v_2-v_3+1)}\,.
\end{align}
For the eigenvalues of the $\CR-$operator corresponding to these
eigenvectors one obtains
\begin{align}\label{ER}
R^{1,n}_{\boldsymbol{\rho}\,\boldsymbol{\sigma}}(\lambda)&=(-1)^n\,f(\lambda+\sigma_2-\rho_2)\,
\frac{\Gamma(n+\sigma_2-\rho_1+1+\lambda)}{\Gamma(n+\rho_2-\sigma_1+1-\lambda)}
\frac{\Gamma(\sigma_3-\rho_2+1+\lambda)}{\Gamma(\rho_3-\sigma_2+1-\lambda)}
\frac{\Gamma(\sigma_3-\rho_1+1+\lambda)}{\Gamma(\rho_3-\sigma_1+1-\lambda)}\,,\\[2mm]
R^{2,m}_{\boldsymbol{\rho}\,\boldsymbol{\sigma}}(\lambda)&=(-1)^m\,f(\lambda+\sigma_2-\rho_2)\,
\frac{\Gamma(m+\sigma_3-\rho_2+1+\lambda)}{\Gamma(m+\rho_3-\sigma_2+1-\lambda)}
\frac{\Gamma(\sigma_3-\rho_1+1+\lambda)}{\Gamma(\rho_3-\sigma_1+1-\lambda)}
\frac{\Gamma(\sigma_2-\rho_1+1+\lambda)}{\Gamma(\rho_2-\sigma_1+1-\lambda)}\,.
\end{align}
We remind that the $\CR-$operator acts on the tensor product
$\VV_1\otimes \VV_2=\VV_{\boldsymbol{\rho}}\otimes\VV_{\boldsymbol{\sigma}}$.
Let us note that with such normalization the eigenvalues of the $\CR-$operator
possess the following properties
\begin{align}\label{RtoR}
R^{1,n}_{\boldsymbol{\rho}\,\boldsymbol{\sigma}}(\lambda)|_{n=\sigma_{12}}=
R^{1,0}_{\boldsymbol{\rho}\,\boldsymbol{\sigma}'}(\lambda)|_{\boldsymbol{\sigma}'=
(\sigma_2,\sigma_1,\sigma_3)}&&\text{and}&&
R^{2,m}_{\boldsymbol{\rho}\,\boldsymbol{\sigma}}(\lambda)|_{m=\sigma_{23}}=
R^{2,0}_{\boldsymbol{\rho}\,\boldsymbol{\sigma}'}(\lambda)|_{\boldsymbol{\sigma}'=
(\sigma_1,\sigma_3,\sigma_2)}\,.
\end{align}
It follows that   the $\CR-$operator satisfies the  relations
\begin{subequations}\label{DD12}
\begin{align}
\CD_{1}^{\sigma_{12}} \CR_{\br\bs}(u)=\CR_{\br\bs'}(u)\CD_{1}^{\sigma_{12}}\,,
 \intertext{when $\sigma_{12}$
 is a
positive integer and $\bs'=P_{12}\bs$ and }
\CD_{2}^{\sigma_{23}} \CR_{\br\bs}(u)=\CR_{\br\bs'}(u) \CD_{2}^{\sigma_{23}}\,,
\end{align}
\end{subequations}
when $\sigma_{23}$  is a
positive integer and $\bs'=P_{23}\bs$. This implies that the coefficients
$r_{12}(u)$ and $r_{23}(u)$ in Eq.~\eqref{twineR} are equal to one.
If both the differences $\sigma_{12}$ and $\sigma_{23}$ are positive
integers, then it follows from
Eqs.~\eqref{DD12} that the coefficient $r_{13}(u)=1$ as well.
This relation,
$r_{13}(u)=1$, remains valid for arbitrary $\sigma_{12},\sigma_{13}$ provided
that their sum, $\sigma_{13}$,  is a positive integer.
To prove this let us represent the operator $\CW_{13}$, Eq.~\eqref{W13} in the polynomial
form
\begin{align}
\CW_{13}=\sum_{k=0}^mC^m_k \frac{\Gamma(r+1)}{\Gamma(r+1-k)} \CD_2^{m-k}\CD_1^{m-k}[\CD_1,\CD_2]^k\,,
\end{align}
where we put $m=\sigma_{13}$ and $r=\sigma_{23}$. This operator maps the lowest weight
vectors $\Phi_k\equiv\Psi_{k,k,m-k}$ (Eq.~\eqref{lwv}) to $1$,
\begin{align}
\CW_{13}\,\Phi_k=c_k(m,r)\cdot 1\,, && c_k(m,r)=m!\,k!\,\frac{\Gamma(r+1)}{\Gamma(r+1-m+k)}\,.
\end{align}
To get the necessary result it is sufficient to verify  that
\begin{align}
\CW_{13}\, \CR_{\br\bs}(u)\,\Phi_0=\CR_{\br\bs'}(u)\,\CW_{13}\,\Phi_0\,,
\end{align}
where $\bs'=(\sigma_3\sigma_2\sigma_1)$. It can be checked by  straightforward calculation.

\section{Appendix: Lowest weight vectors}\label{app:modules}
The equations on the lowest weight vectors in the dual space $\WW$
(see subsect.~\ref{subs:modules})
 take the form
\begin{subequations}\label{EQQ}
\begin{align}
&2\widetilde H_1\Psi=(2x\partial_x+y\partial_y-z\partial_z+m_1)\Psi=\mu_1\Psi\,,\\[2mm]
&2\widetilde H_2\Psi=(2z\partial_z+y\partial_y-x\partial_x+m_2)\Psi=\mu_2\Psi\,,\\[2mm]
\label{12}
&\widetilde
L_{12}\Psi=(x\partial_x^2+\partial_x(y\partial_y-z\partial_z+m_1)+z\partial_y)\Psi=0\,,\\[2mm]
&\widetilde
L_{23}\Psi=(z\partial_z^2+m_2\partial_z-x\partial_y)\Psi=0\,.
\end{align}
\end{subequations}
The first two equations yield
\begin{align}
\Psi(x,y,z)=x^A z^B\, \psi\left(\frac{xz}{y}\right)\,,
\end{align}
where
\begin{align}\label{AB}
&A=\frac13\left(2(\mu_1-m_1)+\mu_2-m_2\right)&\ \ \ \mathrm{and}\ \ \ \ &
&B=\frac13\left(2(\mu_2-m_2)+\mu_1-m_1\right)\,.
\end{align}
The last two equations in \eqref{EQQ} result in the following equations on the function
$\psi(\tau)$, ($\tau=xz/y$)
\begin{subequations}\label{ePhi}
\begin{align}
&\tau^2\psi'(\tau)+\biggl[\tau^2\psi''(\tau)+(2A+2-\mu_1)\,\tau\psi'(\tau)+
A(A-\mu_1+1)\,\psi(\tau)\biggr]=0\,,\\[2mm]
&\tau^2\psi'(\tau)+\biggl[\tau^2\psi''(\tau)+(2B+m_2)\,\tau\psi'(\tau)+B(B-1+m_2)\,\psi(\tau)\biggr]
=0\,.
\end{align}
\end{subequations}
Clearly we are interested in solutions which are series in $1/\tau$.
Provided that $A(A-\mu_1+1)=B(B-1+m_2)=0$  Eqs.~\eqref{ePhi} have the solution
$\psi(\tau)=1$. It gives rise to the four solutions for the function $\Psi(x,y,z)$
\begin{align}
\Psi_1=1\,,&&\Psi_2=x^{1-m_1}\,,&& \Psi_3=z^{1-m_2}\,,&&\Psi_4=x^{1-m_1}\,z^{1-m_2}\,.
\end{align}
The equations~\eqref{ePhi} and \eqref{AB} has another solution:
$B=2-m_1-m_2$, $A=2-m_1-m_2$ or $1-m_1$ and
\begin{align}
\psi_{m_1,m_2}(\tau)={}_2F_0(m_1-1,m_1+m_2-2|1/\tau)\,,
\end{align}
So we get two more solutions for $\Psi(x,y,z)$
\begin{align}
\Psi_5=(xz)^{2-m_1-m_2}\psi_{m_1,m_2}\left(\frac{xz}{y}\right)\,,&&
\Psi_6=x^{1-m_1}z^{2-m_1-m_2}\psi_{m_1,m_2}\left(\frac{xz}{y}\right) \,.
\end{align}
If none of the numbers $1-m_1,1-m_2,2-m_1-m_2$ is a positive integer, then only the vector $\Psi_1$
belong to the dual space $\WW$. If only one of these numbers is positive integer there are
an additional nontrivial lowest weights in the space $\WW$, \
$\Psi_2,\Psi_3$ and $\Psi_5$, respectively.

Let show now that the space $\VV^{(23)}_{\sigma_1\sigma_2\sigma_3}\equiv\VV^{(23)}_{m_1,m_2}$
cannot contain more than one invariant subspace.
Indeed, if the space
$\VV^{(23)}_{\sigma_1\sigma_2\sigma_3}\equiv\VV^{(23)}_{m_1,m_2}$ possesses an invariant
subspace, then there should exist a nontrivial, $\Psi\neq 1$,
lowest weight vector in the dual space $\WW$ with  a nonzero projection on
$\VV^{(23)}_{m_1,m_2}$,
$(\Psi, \Phi)\neq 0$ for some vector $\Phi\in \VV^{(23)}_{m_1,m_2}$. If there exists more
than one invaraint subspace one should find at least two lowest weight vectors with this
property. We remind that space $\VV^{(23)}_{m_1,m_2}$   ($1-m_2$ is a positive integer) is
spanned by the basis vectors $z^k x^n y^p$, where $k\leq -m_2$.
Let us count the number of the nontrivial lowest weight vectors with  a nonzero projection onto
$\VV^{(23)}_{m_1,m_2}$. One easily finds
\begin{itemize}
\item
$1-m_1$ is a positive integer:  There exist five nontrivial lowest weight vectors,
$\Psi_2,\ldots,\Psi_6$, but
only one,
$\Psi_2$, has a nonzero projection.
\item
$2-m_1-m_2$ is a positive integer and $1-m_1$ is not.
There exists two nontrivial lowest weight vectors $\Psi_3$ and $\Psi_5$ and only  last one
has a nonzero projection.
\item
Both $1-m_1$ and $2-m_1-m_2$ are not  positive integers. There are no the lowest weight
vectors with nonzero projection.
\end{itemize}


\end{document}